   \newcommand{\be}[0]{\begin{equation}}
   \newcommand{\ee}[0]{\end{equation}}
   \newcommand{\ba}[0]{\begin{eqnarray}}
   \newcommand{\ea}[0]{\end{eqnarray}}

\documentclass[12pt]{article}
\usepackage{epsfig} \usepackage{amssymb} \usepackage{amsfonts}
\usepackage{epsf,psfrag}
\begin{document}
\Large
\hfill\vbox{\hbox{IPPP/03/09}
            \hbox{DCPT/03/18}}

\nopagebreak

\vspace{0.75cm}
\begin{center}
\LARGE
{\bf All-orders infrared freezing of observables in perturbative QCD}
\vspace{0.6cm}
\Large

D.~M.~Howe\footnote{email:{\tt d.m.howe@durham.ac.uk}} and C.~J.~Maxwell\footnote{email:{\tt c.j.maxwell@durham.ac.uk}}

\vspace{0.4cm}
\large
\begin{em}
Centre for Particle Theory, University of Durham\\
South Road, Durham, DH1 3LE, England
\end{em}

\vspace{1.7cm}

\end{center}
\normalsize
\vspace{0.45cm}

\centerline{\bf Abstract}
\vspace{0.3cm}
We consider a Borel sum definition of all-orders perturbation theory for Minkowskian
QCD observables such as the ${R}_{{e}^{+}{e}^{-}}$ ratio, and show that both this perturbative
component and the additional non-perturbative all-orders Operator Product Expansion (OPE) component
can remain {\it separately} well-defined for all values
of energy $\sqrt{s}$, with the perturbative component dominating as $s\rightarrow{\infty}$,
and with both components contributing
as $s\rightarrow{0}$.
 In the infrared ${s}\rightarrow{0}$ limit the perturbative
correction to the parton model result for ${R}_{{e}^{+}{e}^{-}}$ has an all-orders perturbation theory
component which
smoothly freezes
to the value ${\cal{R}}(0)={2/b}$, where $b=(33-2{N}_{f})/6$ is the first QCD beta-function
coefficient, with ${N}_{f}$ flavours of massless quark. For freezing one requires
${N}_{f}<9$. The freezing behaviour is manifested by the ``Contour-improved'' or
``Analytic Perturbation Theory''(APT), in which an infinite subset of analytical
continuation terms are resummed to all-orders.
We show that for the Euclidean Adler-$D$ function, $D({Q}^{2})$, the perturbative component
remains defined into the infrared if {\it all} the renormalon singularities are taken into
account, but no analogue of the APT reorganisation of perturbation theory is possible.
We perform phenomenological comparisons of suitably smeared low-energy data for
the ${R}_{{e}^{+}{e}^{-}}$  ratio, with the perturbative freezing predictions,
and find good agreement.
\newpage
\section*{1 Introduction}
In this paper we wish to address the question of whether QCD perturbation theory
can be used to make predictions in the low-energy infrared regime where one
expects non-perturbative effects to dominate. Such an extension of the applicability
of perturbation theory, beyond the ultraviolet regime of Asymptotic Freedom, would
obviously enable one to test QCD in new ways. Reorganisations of fixed-order perturbation
theory exhibiting a stable infrared freezing behaviour have previously been formulated
and studied, these include the so-called ``Analytic Perturbation Theory'' (APT) approach
initiated by Shirkov and Solovtsov in Refs.\cite{apt1}  (for a review see Ref.\cite{apt2a})
, and the Variational Perturbation
Theory (VPT) approach \cite{vpt1}. Our discussion will address the more fundamental
question of how all-orders QCD perturbation theory, and the non-perturbative
Operator Product Expansion (OPE) contribution, can remain defined in the low-energy
regime. We will discover that this is possible for Minkowskian observables, and that
the APT approach should be asymptotic to the all-orders perturbative result which
also exhibits the same infrared freezing behaviour found with APT. For Euclidean
quantities, however, we will find that
 all-orders perturbation theory only exhibits stable infrared
behaviour if one has complete information on the perturbative corrections to all-orders,
and that this behaviour is not in general related to the infrared behaviour found using
APT.

We will focus our discussion on
the ${R}_{{e}^{+}{e}^{-}}(s)$ ratio, at c.m. energy $\sqrt{s}$. This is a Minkowskian
quantity derived by analytical continuation from the Euclidean QCD vacuum
polarization function. The corrections to the parton model result for ${R}_{{e}^{+}{e}^{-}}(s)$
will consist of a perturbative part, which can be developed as a power series in the
renormalized QCD coupling ${a}(s){\equiv}{\alpha}_{s}(s)/{\pi}$, and a non-perturbative
part which can be developed as an OPE in powers of ${\Lambda}^{2}/s$,
the first term corresponding to the lowest dimension relevant operator, the gluon
condensate, being proportional to ${({\Lambda}^{2}/s)}^{2}$. The key point is that
the combination of the all-orders perturbation series and OPE must be well-defined
at all values of $s$, since ${R}_{{e}^{+}{e}^{-}}(s)$ is a physical quantity. Each
part by itself, however, exhibits pathologies. Specifically, the perturbation series
exhibits $n!$ growth in the perturbative coefficients, at large-orders $n$.
 Attempts to define the all-orders
sum of the perturbation series using a Borel integral run into the difficulty that
there are singularities on the integration contour termed infrared renormalons \cite{r1}.
It turns out, however, that the resulting ambiguity in defining the Borel integral
is of the same form as ambiguities in the coefficient functions involved in the OPE,
and so choosing a particular regulation of the Borel integral (such as principal value)
induces a corresponding definition of the coefficient functions, and the sum of
the two components is well-defined \cite{r1,r2}.
 There is a further crucial pathology of the
Borel integral, which we shall refer to as the ``Landau divergence''. This means
that at a critical energy ${s}={s}_{L}$, the Borel integral diverges. It should be
stressed that the value of ${s}_{L}$ should not be confused with the ``Landau pole''
or ``Landau ghost'' in the QCD coupling $a(s)$. The ``Landau ghost'' is
completely unphysical and scheme-dependent, whereas the divergence of the Borel
integral is completely scheme-independent \cite{r1}. For Minkowskian quantities
such as ${R}_{{e}^{+}{e}^{-}}$ there is an oscillatory factor in the Borel transform in the
integrand, arising from the analytical continuation from Euclidean to Minkowskian,
 which means that the Borel integral is finite at ${s}={s}_{L}$, and
diverges for $s<{s}_{L}$.
 To go to lower energies than ${s}_{L}$ we shall show that
 one needs to modify
the form of the Borel integral, the modified form now having singularities
on the integration contour corresponding to {\it ultraviolet} renormalons, correspondingly
to go below $s={s}_{L}$ one needs to resum the OPE to all-orders and recast it as
a modified expansion in powers of ${s}/{\Lambda}^{2}$. One then finds that the
ambiguities in regulating this modified Borel integral, are of the same form as
ones in the modified OPE, and for $s<{s}_{L}$ the sum of the two components is
again well-defined. In the infrared ${s}\rightarrow{0}$ limit the modified OPE resulting
from resummation can contain a constant term independent of $s$ even though such a term
is not invisible in perturbation theory, so both the perturbative and non-perturbative
components will contribute to the infrared freezing limit.
 The oscillatory factor
in the Borel integral means that it freezes smoothly to $2/b$ in the infrared,
where $b=(33-2{N}_{f})/6$ is the first SU($3$) QCD beta-function coefficient, with
${N}_{f}$ quark flavours. \\

The arguments sketched above suggest that the all-orders perturbative and non-perturbative
components for Minkowskian quantities such as ${R}_{{e}^{+}{e}^{-}}(s)$ can
{\it separately} remain defined at all energies, with the perturbative part being
dominant in the ultraviolet and both components contributing in the
infrared limit. One can then compare
all-orders perturbative predictions with data, having suitably smeared and averaged
over resonances \cite{r3} to suppress the non-perturbative mass threshold effects.
 In practice,
of course, we do not have exact all-orders perturbative information. We know exactly
the perturbative coefficients of the corrections to the parton model result for
${R}_{{e}^{+}{e}^{-}}$ to next-next-leading order (NNLO), i.e. including terms
of order ${\alpha}_{s}^{3}$ \cite{r4}. Clearly, conventional fixed-order perturbation
theory for ${R}_{{e^+}{e^-}}$ will not exhibit the freezing behaviour in the
infra-red to be expected for the all-orders perturbation theory. What is required is
a rearrangement of fixed-order perturbation theory which has freezing behaviour in
the infrared. As we have discussed in a recent paper \cite{r5} the resummation to
all-orders of the convergent subset of analytical continuation terms (``large-${\pi}^{2}$'' terms)
, arising when the perturbative corrections to the Euclidean Adler-$D(-s)$ function
at a given order are continued to the Minkowskian ${R}_{{e}^{+}{e}^{-}}(s)$, recasts the
perturbation series as an expansion in a set of functions ${A}_{n}(s)$ which are
well-defined for all values of $s$, vanishing as ${s}\rightarrow{\infty}$ in accord with Asymptotic Freedom,
 and with all but ${A}_{1}(s)$ vanishing in the infrared limit,
with ${A}_{1}(s)$ approaching $2/b$ to provide infrared freezing behaviour to all-orders
in perturbation theory. This ``contour-improved'' perturbation theory (CIPT) approach
is equivalent to the Analytic Perturbation Theory (APT) mentioned above \cite{apt1}
in the case of ${R}_{{e}^{+}{e}^{-}}(s)$. We gave explicit expressions for
the functions ${A}_{n}(s)$. At the two-loop level these can be expressed in terms of
the Lambert $W$-function \cite{r7,r8}. To make contact with the all-orders perturbative
result represented as a Borel integral, we note that the CIPT/APT reorganisation of
perturbation theory corresponds to leaving the oscillatory factor in the Borel
transform intact whilst expanding the remaining factor as a power series. Integrating
term-by-term then yields the functions ${A}_{n}(s)$. The presence of the oscillatory
factor in these integrals guarantees that the ${A}_{n}(s)$ are well-defined at all
energies. The CIPT/APT series should thus be asymptotic to the Borel integral at
both ultraviolet and infrared energies. Whilst a reorganised fixed-order perturbation series
exhibiting stable infrared freezing behaviour is possible for Minkowskian quantities,
we shall show that it is not possible for Euclidean observables such as the Adler $D$-function,
$D({Q}^{2})$. For Euclidean observables the Borel integral does not contain the
oscillatory factor and so is potentially divergent at $s={s}_{L}$, although as we shall
show in the so-called leading-$b$ approximation \cite{r1,r2}, the divergence is cancelled
if {\it all} the infrared and ultraviolet renormalon singularities are included, and once
again perturbative and non-perturbative components which are separately
well-defined at all energies can be obtained. This is only possible in the leading-$b$
approximation, however. \\

The plan of the paper is as follows. In Section 2 we shall describe the CIPT/APT
reorganisation of fixed-order perturbation theory for ${R}_{{e}^{+}{e}^{-}}$, reviewing
the results of Ref.\cite{r5}. In Section 3 we consider how for Minkowskian observables
one can define all-orders perturbation theory, and the all-orders non-perturbative OPE
in such a way that each component remains well-defined at all energies. The link between
the all-orders perturbative result and the reorganised CIPT/APT fixed-order perturbation
theory is emphasised. We then briefly consider the corresponding problem for Euclidean observables.
 In Section 4 we perform some phenomenological studies in which we compare low
energy experimental data for ${R}_{{e}^{+}{e}^{-}}(s)$ with the CIPT/APT perturbative
predictions. Section 5 contains a Discussion and Conclusions. \\

\section*{2 Infra-red freezing of ${R}_{{e}^{+}{e}^{-}}$- CIPT/APT}

We begin by defining the ${R}_{{e^+}{e^-}}$ ratio at c.m. energy $\sqrt{s}$,
\be
{R}_{{e^+}{e^-}}(s){\equiv}\frac{{\sigma}
_{tot}({e}^{+}{e}^{-}\rightarrow\rm{hadrons})}{{\sigma}(
{e}^{+}{e}^{-}\rightarrow{\mu}^{+}{\mu}^{-})}=3{\sum_{f}}{Q}_{f}^{2}(1+{\cal{R}}(s))\;.
\ee
Here the $Q_f$ denote the electric charges of the different flavours of quarks, and ${\cal{R}}(s)$
denotes the perturbative corrections to the parton model result, and has a perturbation series
of the form,
\be
{\cal{R}}(s)=a+
{\sum_{n>0}}{r}_{n}{a}^{n+1}\;.
\ee
Here $a{\equiv}{\alpha}_{s}({\mu}^{2}
)/{\pi}$ is the renormalized coupling, and the
coefficients $r_1$ and $r_2$ have been computed in the $\overline{MS}$ scheme with
renormalization scale ${\mu}^{2}=s$ \cite{r4}. We can consider the $s$-dependence of
${\cal{R}}(s)$ at NNLO,
\be
s\frac{d{\cal{R}}(s)}{ds}=-\frac{b}{2}{\rho}({\cal{R}}){\equiv}-\frac{b}{2}{\cal{R}}^{2}(1+c{\cal{R}}+{\rho}_{2}
{\cal{R}}^{2})\;.
\ee
Here $c=(153-19{N}_{f})/12b$ is the second universal QCD beta-function coefficient, and
${\rho}_{2}$ is the NNLO effective charge beta-function coefficient \cite{r11}, an RS-invariant
combination of ${r}_{1},{r}_{2}$ and beta-function coefficients. The condition for ${\cal{R}}(s)$ to
approach the infrared limit ${\cal{R}}^{\ast}$ as $s\rightarrow{0}$ is for the Effective Charge
beta-function to have a non-trivial zero, ${\rho}({\cal{R}}^{\ast})=0$. At NNLO the condition
for such a zero is ${\rho}_{2}<0$. Putting ${N}_{f}=2$ active flavours we find for the NNLO
RS invariant, ${\rho}_{2}=-9.72$, so that ${\cal{R}}(s)$ apparently freezes in the infrared to
${\cal{R}}^{\ast}=0.43$. The freezing behaviour was first investigated in a pioneering paper by
Mattingly and Stevenson \cite{r12} in the context of the Principle of Minimal Sensitivity (PMS)
approach. However, it is not obvious that we should believe this apparent NNLO freezing
result \cite{r13}. In fact ${\rho}_{2}$ is dominated by a large ${b}^{2}{\pi}^{2}$ term arising from
analytical continuation (AC) of the Euclidean Adler $D(-s)$ function to the Minkowskian
$R(s)$, with ${\rho}_{2}=9.40-{\pi}^{2}{b}^{2}/12$. Similarly the ${\rm{N}}^{3}$LO invariant
${\rho}_{3}$ will contain the large AC term $-5c{\pi}^{2}{b}^{2}/12$. This suggests that in
order to check freezing we need to resum the AC terms to {\it all-orders}.

${R}_{{e}^{+}{e}^{-}}$ is directly related
to the transverse part of the correlator of two vector currents in the Euclidean region,
\be
({q}_{\mu}{q}_{\nu}-{g}_{{\mu}{\nu}}{q}^{2}){\Pi}(s)=4{\pi}^{2}i{\int}{d}^{4}x{e}^{iq.x}<0|T[{j}_
{\mu}(x){j}_{\nu}(0)]|0>\;,
\ee
where $s=-{q}^{2}>0$. To avoid an unspecified constant it is convenient to take a
logarithmic derivative with respect to $s$ and define the Adler $D$-function,
\be
{D}(s)=
-s\frac{d}{ds}{\Pi}(s)\;.
\ee
This can be represented by Eq.(1) with the perturbative corrections ${\cal{R}}(s)$ replaced by
\be
{\cal{D}}(s)=a+
{\sum_{n>0}}{d}_{n}{a}^{n+1}\;.
\ee
 The Minkowskian observable ${\cal{R}}(s)$ is related to
${\cal{D}}(-s)$ by analytical continuation from Euclidean to Minkowskian. One may write
the dispersion relation,
\be
{\cal{R}}(s)=\frac{1}{2{\pi}i}{\int_{-s-i\epsilon}^{-s+i\epsilon}}{dt}\frac{{\cal{D}}(t)}{t}\;.
\ee
Written in this form it is clear that the ``Landau pole'' in the coupling $a(s)$, which
lies on the positive real $s$-axis,
is not a problem, and ${\cal{R}}(s)$ will be
defined for all $s$.
 The dispersion relation can be
reformulated as an integration around a circular contour in the complex
energy-squared $s$-plane \cite{r14,r15},
\be
{\cal{R}}(s)=\frac{1}{2{\pi}}{\int_{-{\pi}}^{\pi}}{d{\theta}}{\cal{D}}(s{e}^{i{\theta}})\;.
\ee
One should note, however, that this is only equivalent to the dispersion relation of Eq.(7)
for values of $s$ above the
``Landau pole''.
Expanding ${\cal{D}}(s{e}^{i{\theta}})$ as a power series in $\bar{a}{\equiv}{a}(s{e}^{i{\theta}})$,
and performing the ${\theta}$ integration term-by-term, leads to a ``contour-improved''
perturbation series, in which at each order an infinite subset of analytical continuation
terms present in the conventional perturbation series of Eq.(2) are resummed. It is this
complete analytical continuation that builds the
 freezing of ${\cal{R}}(s)$. We shall begin by considering the ``contour-improved'' series for
the simplified case of a one-loop coupling. The one-loop coupling will be given by
\be
a(s)=\frac{2}
{b{\ln}(s/{\tilde{\Lambda}}_{\overline{MS}}^{2})}\;.
\ee
As described above one can then obtain the ``contour-improved'' perturbation series
for ${\cal{R}}(s)$,
\be
{\cal{R}}(s)={A}_{1}(s)+{\sum_{n=1}^{\infty}}{d}_{n}{A}_{n+1}(s)\;,
\ee
where the functions ${A}_{n}(s)$ are defined by,
\ba
{A}_{n}(s)&{\equiv}&\frac{1}{2{\pi}}{\int_{-{\pi}}^{\pi}}{d{\theta}}{\bar{a}}^{n}=
\frac{1}{2{\pi}}{\int_{-\pi}^{\pi}}
{d{\theta}}\frac{{a}^{n}(s)}{{[1+ib{\theta}a(s)/2
]}^{n}}\;.
\ea
This is an elementary integral which can be evaluated in closed-form as \cite{r5}
\ba
{A}_{1}(s)&=&\frac{2}{{\pi}b}{\arctan}\left(\frac{{\pi}ba(s)}{2}\right)
\nonumber \\
{A}_{n}(s)&=&
\frac{2{a}^{n-1}(s)}{b{\pi}(1-n)}
Im\left[{\left(1+
\frac{ib{\pi}a(s)}{2}\right)}^{1-n}
\right]\;(n>1)\;.
\ea
We then obtain the one-loop ``contour-improved'' series for ${\cal{R}}(s)$,
\be
{\cal{R}}(s)=\frac{2}{\pi{b}}{\arctan}\left(\frac{\pi{b}{a}(s)}{2}\right)+
{d}_{1}\left[\frac{{a}^{2}(s)}{(1+{b}^{2}{\pi}^{2}{a}^{2}(s)/4)}\right]
+{d}_{2}\left[\frac{{a}^{3}(s)}{{(1+{b}^{2}{\pi}^{2}{a}^{2}(s)/4)}^{2}}\right]+\ldots\;.
\ee

The first $\arctan$ term is well-known, and corresponds to resumming the infinite
subset of analytical continuation terms in the standard perturbation series of
Eq.(2) which are independent of the $d_n$ coefficients. Subsequent terms
corrrespond to resumming to all-orders the infinite subset of terms in Eq.(2)
proportional to ${d}_{1},{d}_{2},{\ldots}$, etc. In
 each case the resummation is {\it convergent}, provided that $|a(s)|<2/{\pi}{b}$.
 In the ultraviolet $s\rightarrow{\infty}$ limit
the ${A}_{n}(s)$ vanish as required by asymptotic freedom. In the infrared
$s\rightarrow{0}$ limit, the one-loop coupling $a(s)$ has a ``Landau'' singularity
at $s={\tilde{\Lambda}}_{\overline{MS}}^{2}$.
 However, the functions ${A}_{n}(s)$
resulting from resummation, if analytically continued,
 are well-defined for all real values of $s$. ${A}_{1}(s)$
smoothly approaches from below the asymptotic infrared value $2/b$, whilst for
$n>1$ the ${A}_{n}(s)$ vanish.
 Thus, as claimed,
 ${\cal{R}}(s)$ is asymptotic to
$2/b$ to all-orders in perturbation theory.
 We postpone the crucial question of how
to define all-orders perturbation theory in the infrared region until the next
Section. We should also note that the functions ${A}_{n}(s)$ in Eq.(12) can also
be obtained by simple manipulation of the dispersion relation in Eq.(7), which
is defined for all real $s$. This avoids the possible objection that the contour integral
in Eq.(8) is only defined for $s$ above the ``Landau pole''.\\

Beyond the simple one-loop approximation the freezing is most easily analysed by
choosing a renormalization scheme in which the
beta-function equation has its two-loop form,
\be
\frac{\partial{a}({\mu}^{2})}
{\partial{\ln}{\mu}^{2}
}=-\frac{b}{2}
{a}^{2}({\mu}^{2})(1+c{a}({\mu}^{2}))\;.
\ee
This corresponds to a so-called 't Hooft scheme \cite{r16} in which the non-universal
beta-function coefficients are all zero. Here $c=(153-19{N_f})/12b$ is the
second universal beta-function coefficient. For our purposes the
 key feature of these schemes is
that the coupling can be expressed analytically in closed-form in terms of the
Lambert $W$ function
, defined implicitly by $W(z){\exp}(W(z))=z$ \cite{r17,r18}. One has
\ba
{a}({\mu}^{2})
&=&-\frac{1}{c[1+{W}_{-1}(z(\mu))]}
\nonumber \\
z(\mu)&{\equiv}&-\frac{1}{e}{\left(\frac{\mu}{{\tilde{\Lambda}}_{\overline{MS}}}\right)}^{-b/c}\;,
\ea
where ${\tilde{\Lambda}}_{\overline{MS}}$ is defined according to the convention of \cite{r19}
, and is related to the standard definition \cite{r20} by ${\tilde{\Lambda}}_{\overline{MS}}=
{(2c/b)}^{-c/b}
{\Lambda}_{\overline{MS}}$. The ``$-1$'' subscript on $W$ denotes the
branch of the Lambert $W$ function required for Asymptotic Freedom, the nomenclature being that of Ref.\cite{r18}. Assuming
a choice
of renormalization scale ${\mu}^{2}=xs$ , where $x$ is a dimensionless constant, for
the perturbation series of ${\cal{D}}(s)$ in Eq.(5), one can then expand
the integrand in Eq.(6) for ${\cal{R}}(s)$ in powers of ${\bar{a}}\equiv{a}(xs{e}^{i{\theta}})$
, which can be expressed in terms of the Lambert $W$ function using Eq.(15),
\be
{\bar{a}}=\frac{-1}{c[1+W(A(s){e}^{iK{\theta}})]}
\ee
where
\be
A(s)=\frac{-1}{e}{\left(\frac{\sqrt{xs}}{{\tilde{\Lambda}}_{\overline{MS}}}\right)}^{-{b/c}}\;\;\;,
{K}=\frac{-b}{2c}\;.
\ee
The functions ${A}_{n}(s)$ in the ``contour-improved'' series
 are then given, using Eqs(15,16), by
\ba
{A}_{n}(s)&{\equiv}&\frac{1}{2{\pi}}{\int_{-{\pi}}^{\pi}}{d{\theta}}{\bar{a}}^{n}=
\frac{1}{2{\pi}}{\int_{-\pi}^{0}}{d{\theta}}\frac{{(-1)}^{n}}{c^n}{[1+{W}_{1}(A(s){e}^{iK{\theta}})]}^{-n}
\nonumber \\
&+&\frac{1}{2\pi}{\int_{0}^{\pi}}{d{\theta}}\frac{{(-1)}^{n}}{c^n}{[1+{W}_{-1}(A(s){e}^{iK{\theta}})]}^{-n}\;.
\ea
Here the appropriate branches of the $W$ function are used in the two regions of integration. As discussed
in Refs.\cite{r7,r8},
by making the change of variable $w=W(A(s){e}^{iK{\theta}})$ we can then obtain
\be
{A}_{n}(s)=\frac{{(-1)}^{n}}{2iK{c^n}{\pi}}{\int^{{W}_{-1}(A(s){e}^{iK\pi})}_{{W}_{1}(A(s){e}^{-iK\pi})}}
\frac{dw}{w{(1+w)}^{n-1}}\;.
\ee
Noting
 that ${W}_{1}(A(s){e}^{-iK\pi})={[{W}_{-1}(A(s){e}^{iK\pi})]}^{\ast}$, we can evaluate the elementary integral to
obtain for $n=1$,
\be
A_{1}(s)=\frac{2}{b}-\frac{1}{\pi{K}c}Im[{\ln}({W}_{-1}(A(s){e}^{iK\pi}))]\;,
\ee
where the $2/b$ term is the residue of the pole at $w=0$. For $n>1$ we obtain
\be
{A}_{n}(s)=\frac{{(-1)}^{n}}{{c^n}K{\pi}}Im\left[{\ln}\left(\frac{{W}_{-1}(A(s){e}^{iK\pi})}{1+{W}_{-1}(A(s){e}^{iK\pi})}\right)
+{\sum_{k=1}^{n-2}}\frac{1}{k{(1+{W}_{-1}(A(s){e}^{iK\pi}))}^k}\right]\;.
\ee
Crucially the contribution from the poles at $w=0$ and $w=-1$ cancel exactly.
Equivalent expressions have been
obtained in the APT approach \cite{r8}.
 Provided that
${b/c}>0$, which will be true for ${N_f}<9$, we find the same behaviour as in the one-loop
case with the ${A}_{n}(s)$ vanishing in the ultraviolet limit consistent with Asymptotic
Freedom, and with ${A}_{n}(s)$ vanishing in the infrared limit for $n>1$, and ${A}_{1}(s)$
freezing to $2/b$. To the extent that the freezing holds to all-orders in perturbation
theory it should hold irrespective of the choice of renormalization scheme (RS), The
use of the 't Hooft scheme simply serves to make the freezing manifest.
 In Figures 1-3 we plot the functions ${A}_{1}(s),{A}_{2}(s)$ and ${A}_{3}(s)$, respectively,
as functions of $(
sx/{\tilde{\Lambda}}_{\overline{MS}}^{2})$.
\nopagebreak
\begin{figure}
\begin{center}
\epsfig{file=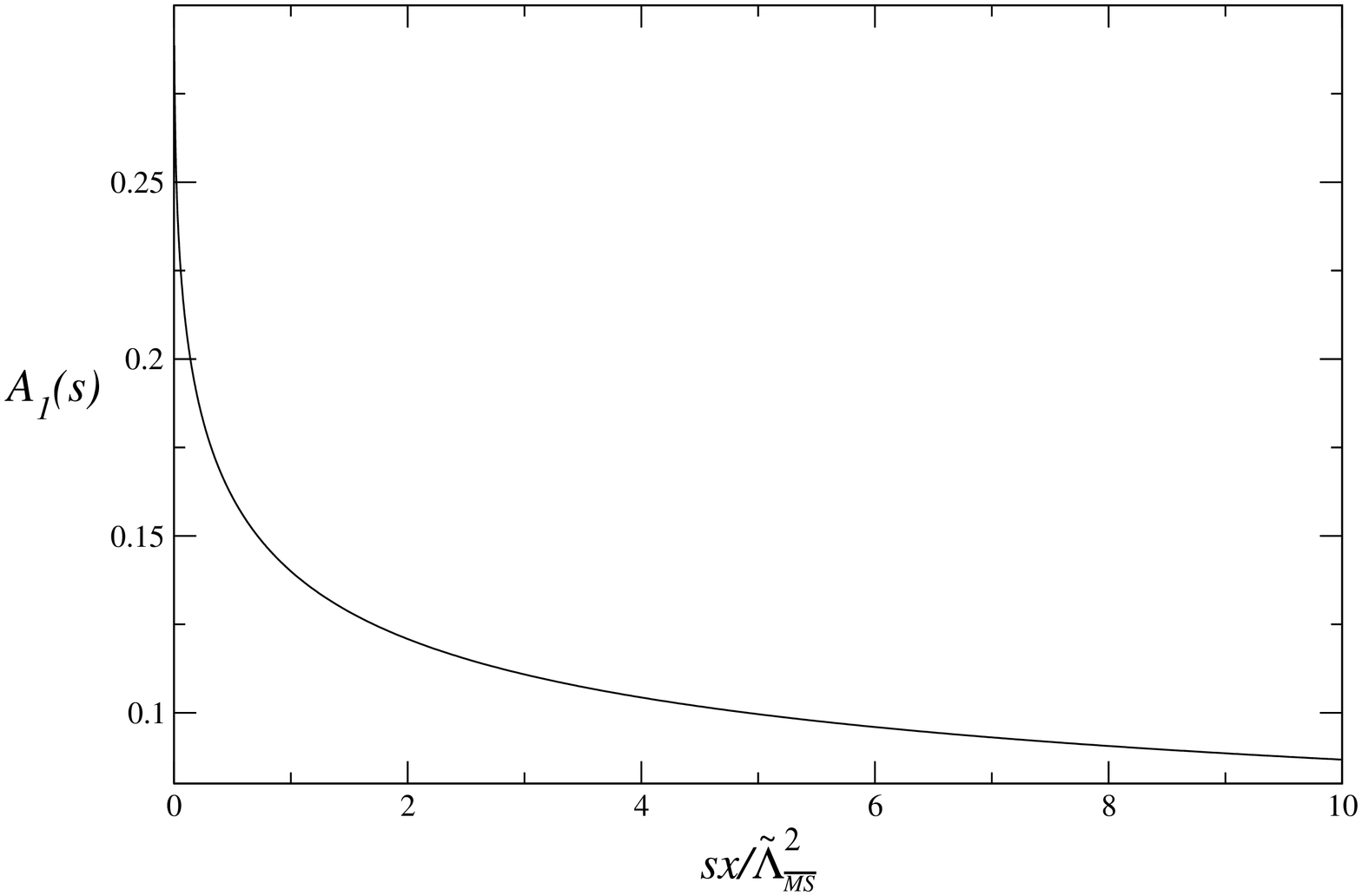,angle=270,width=15cm}
\caption{The function ${A}_{1}(s)$ of Eq.(20)
versus $sx/{\tilde{\Lambda}}_{\overline{MS}}^{2}$. We assume
${N_f}=2$ flavours of quark.}

\label{fig:f1}
\end{center}
\end{figure}
\begin{figure}
\begin{center}
\epsfig{file=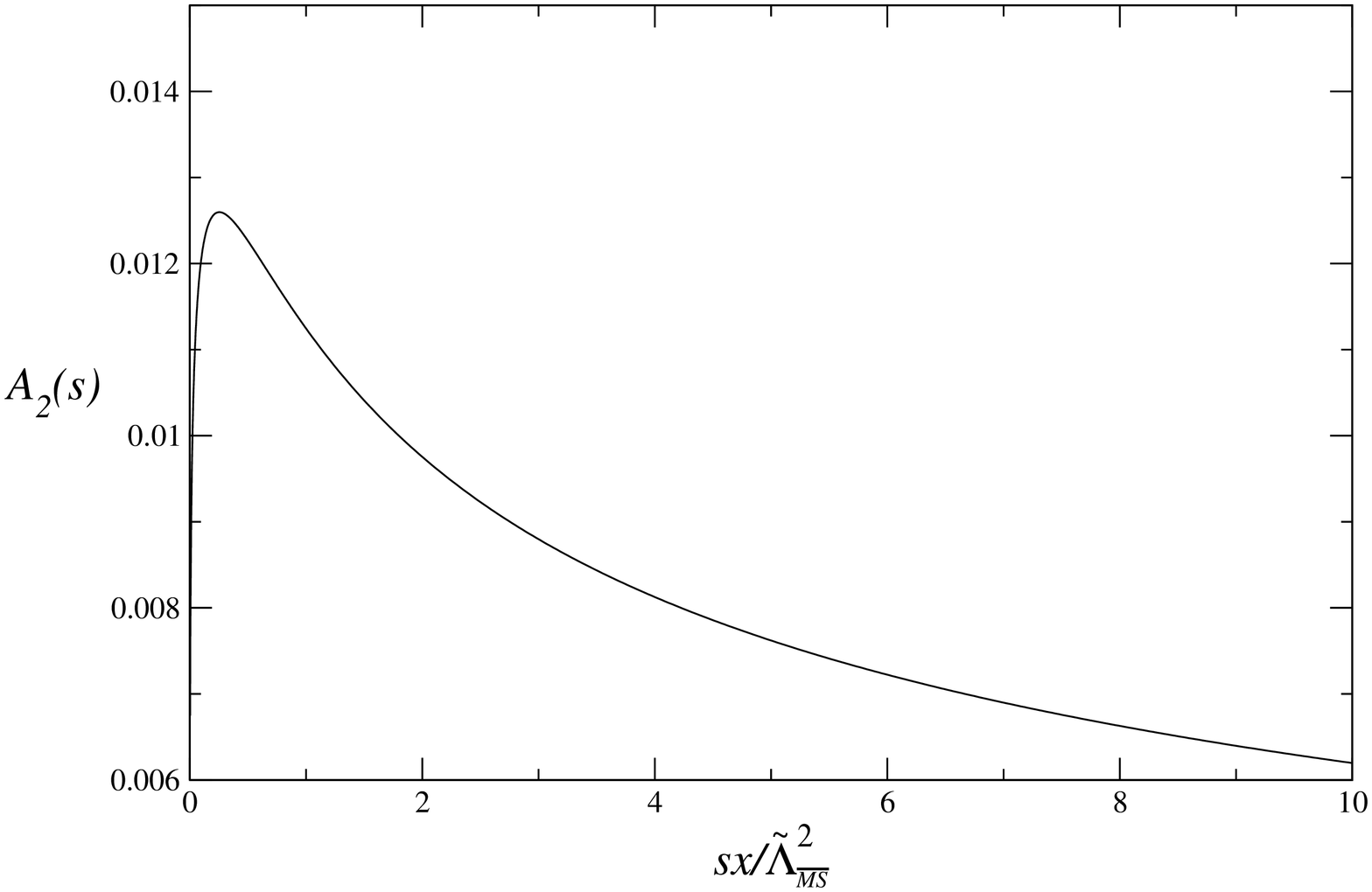,angle=270,width=15cm}
\caption{As Fig.1 but for ${A}_{2}(s)$ of Eq.(21).}
\label{fig:f2}
\end{center}
\end{figure}
\newpage
\begin{figure}
\begin{center}
\epsfig{file=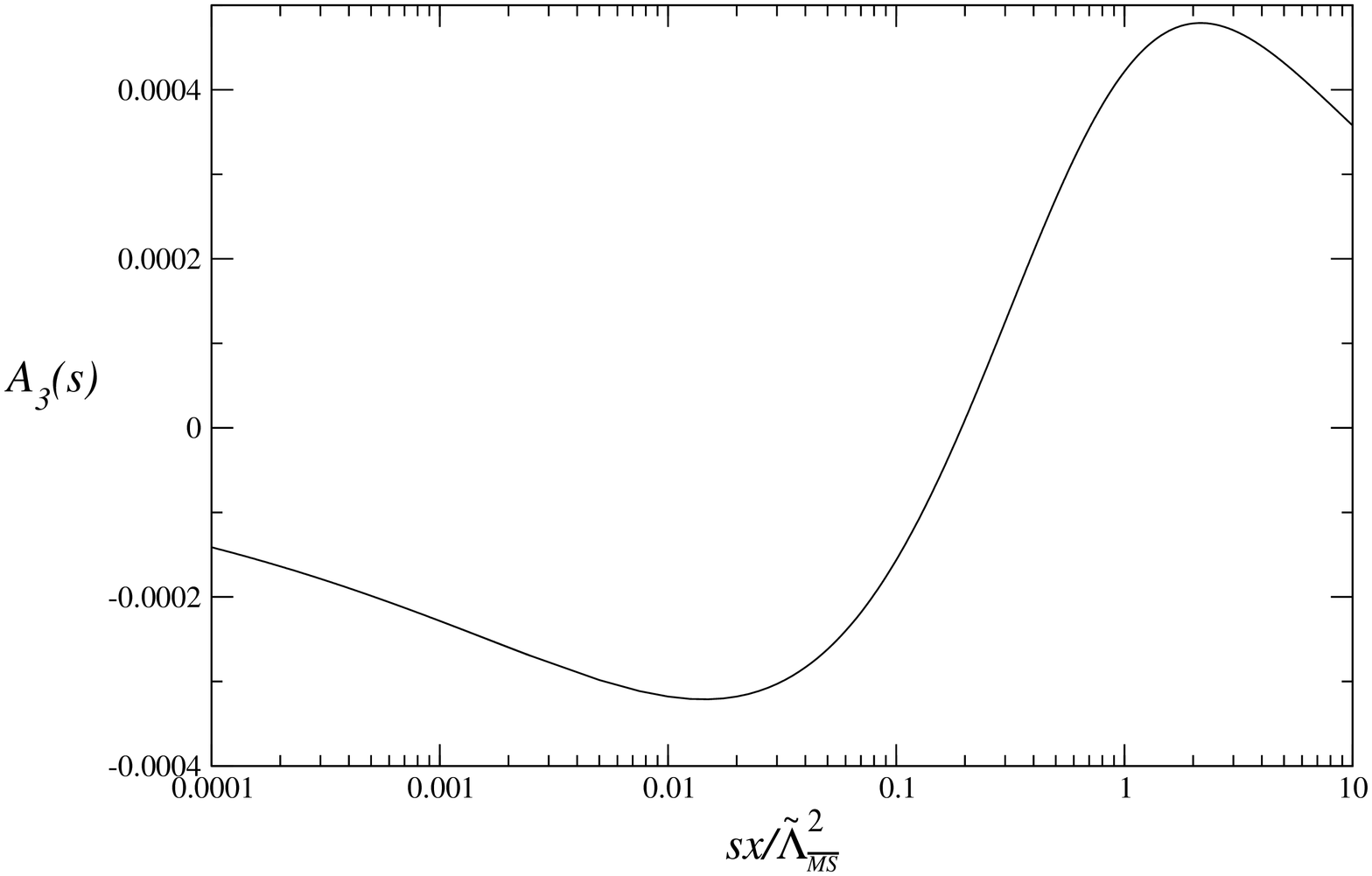,angle=270,width=15cm}
\caption{As Fig.1 but for ${A}_{3}(s)$ of Eq.(21).}
\label{fig:f3}
\end{center}
\end{figure}
Having shown how fixed-order perturbation theory can be reorganised so
that it exhibits well-behaved freezing behaviour in the infra-red, we turn
in the next section to a discussion of how all-orders perturbation theory
and the all-orders non-perturbative OPE, can be defined in such a way that
they remain well-defined at all energies.
\section*{3 All-orders perturbation theory and OPE}
The corrections to the Adler $D$ function, ${\cal{D}}({Q}^{2})$, can
be split into a perturbative part, ${\cal{D}}_{PT}({Q}^{2})$,
 and a non-perturbative Operator Product Expansion (OPE) part, ${\cal{D}}_{NP}({Q}^{2})$,
\be
{\cal{D}}({Q}^{2})={\cal{D}}_{PT}({Q}^{2})+{\cal{D}}_{NP}({Q}^{2})\;.
\ee
The PT component is formally just the resummed perturbation series of Eq.(6),
\be
{\cal{D}}_{PT}({Q}^{2})={a}({Q}^{2})+{\sum_{n>0}}{d}_{n}{a}^{n+1}({Q}^{2})\;.
\ee
In addition one has the non-perturbative OPE contribution,
\be
{\cal{D}}_{NP}({Q}^{2})={\sum_{n}}\frac{{C}_{n}({Q}^{2},{\mu}^{2})<{\cal{O}}_{n}({\mu}^{2})>}{{Q}^{2n}}\;,
\ee
where the sum is over the relevant operators ${\cal{O}}_{n}$ of dimension $2n$. $\mu$ denotes the factorization
scale, and $C_n$ is the coefficient function.
 For the Adler $D$ function
the lowest dimension relevant operator is the dimension four gluon condensate,
\be
<0|{G}^{a}_{{\mu}{\nu}}{G}_{a}^{{\mu}{\nu}}|0>\;.
\ee
It will be convenient to scale out the dimensionful factor ${\tilde{\Lambda}}^{2n}$ from the
operator expectation value, and combine it with the coefficient function to obtain the overall coefficient
${\cal{C}}_{n}({Q}^{2},{\mu}^{2})$. We can then
write the ${\cal{D}}_{NP}({Q}^{2})$ component in the form,
\be
{\cal{D}}_{NP}({Q}^{2})={\sum_{n}}{\cal{C}}_{n}{\left(\frac{{\tilde{\Lambda}}^{2}}{Q^2}\right)}^{n}\;.
\ee
We have suppressed the ${\mu}^{2}$ and $Q^2$ dependence of the coefficient ${\cal{C}}_{n}$. The
coefficients are themselves series expansions in $a$.
\be
{\cal{C}}_{n}=K{a}^{{\delta}_{n}}({\mu}^{2})[1+O(a)]\;.
\ee
Here $K$ is an undetermined non-perturbative normalisation, and ${\delta}_{n}$ is related to the anomalous
dimension of the operator concerned.

The definition of the all-orders perturbative component in Eq.(23) needs care. The series has
zero radius of convergence in the coupling $a$. A direct way of seeing this is to consider
the large-${N}_{f}$ expansion of the perturbative coefficient $d_n$,
\be
{d}_{n}={d}_{n}^{[n]}{N}_{f}^{n}+{d}_{n}^{[n-1]}{N}_{f}^{n-1}+\ldots+{d}_{n}^{[0]}\;.
\ee
The leading large-${N}_{f}$ coefficient, ${d}_{n}^{[n]}$, can be computed exactly to all-orders
since it derives from a restricted set of diagrams in which a chain of $n$ fermion bubbles (renormalon chain)
is inserted in the initiating quark loop. Working in the so-called $V$-scheme
, which corresponds to $\overline{MS}$ subtraction with scale ${\mu}^{2}={e}^{-5/3}{Q}^{2}$, one
finds the exact large-${N}_{f}$ result \cite{r20a,r21,r22},
\ba
{d}_{n}^{[n]}(V)&=&\frac{-2}{3}(n+1){\left(\frac{-1}{6}\right)}^{n}\left[-2n-\frac{n+6}{{2}^{n+2}}\right.
  \nonumber \\
&+&\left.\frac{16}{n+1}{\sum_{\frac{n}{2}+1>m>0}}m(1-{2}^{-2m})(1-{2}^{2m-n-2}){\zeta}_{2m+1}\right]{n}!\;.
\ea
The ${n}!$ growth of coefficients means that the perturbation series is at best an
asymptotic one. To arrive at a function to which it is asymptotic one can use a
Borel integral representation, writing
\be
{\cal{D}}_{PT}({Q}^{2})={\int_{0}^{\infty}}{dz}{e}^{-z/a({Q}^{2})}B[{\cal{D}}](z)\;.
\ee
Here $B[{\cal{D}}](z)$ is the Borel transform, defined by,
\be
B[{\cal{D}}(z)]={\sum_{n=0}^{\infty}}\frac{{z}^{n}{d}_{n}}{n!}\;.
\ee
On performing the Borel integral term-by-term one reconstructs the divergent formal
perturbation series for ${\cal{D}}_{PT}$. If the series for the Borel transform has
finite radius of convergence, by analytical continuation to the whole region
of integration one can then define the Borel Sum, provided that the Borel integral
exists. On general grounds \cite{r23,r24} one expects that in renormalisable field theories
the Borel transform will contain branch point singularities on the real axis in
the complex $z$ plane, at positions ${z}={z}_{k}{\equiv}\frac{2k}{b}$ corresponding to
infrared renormalons, ${IR}_{k}$, $k=1,2,3,\ldots$, and at $z=-{z}_{k}$ corresponding
to so-called ultraviolet renormalons, ${UV}_{k}$. Here $b$ is the first beta-function coefficient,
so that for QED with $N_f$ fermion species $b=-\frac{2}{3}{N}_{f}$, whilst for $\rm{SU}(3)$ QCD with
${N}_{f}$ active quark flavours, $b=(33-2{N}_{f})/6$. Thus in QED there are ultraviolet
renormalon singularities on the positive real axis, and hence the Borel integral will be
ambiguous. In QCD with ${N}_{f}<33/2$ flavours, so that the theory is asymptotically free, and $b>0$,
 there are infrared renormalon singularities on the positive real axis
making the Borel integral again ambiguous. For both field theories all-orders perturbation
theory by itself is not sufficient. The presence of singularities on the integration
contour means that there is an ambiguity depending on whether the contour is
taken above or below each singularity. It is easy to check that, taking ${\cal{D}}$ in the
Borel integral of Eq.(30) to be a generic QED or QCD observable with branch point singularities
${(1-z/|{z}_{k}|)}^{-{\gamma}_{k}}$ in the Borel plane, the resulting ambiguity for the singularity
at $z=|{z}_{k}|$ is of the form
\be
{\Delta}{\cal{D}}_{PT}\sim{K}{e}^{-|{z}_{k}|/{a}({Q}^{2})}{a}^{1-{\gamma}_{k}}\;,
\ee
where $K$ is complex. Using the one-loop form for the coupling,
${a}({Q}^{2})=2/b{\ln}({Q}^{2}/{\tilde{\Lambda}}^{2})$, one finds that in the QCD case,
\be
{\Delta}{\cal{D}}_{PT}{\approx}K{a}^{1-{\gamma}_{k}}{\left(\frac{{\tilde{\Lambda}}^{2}}{{Q}^{2}}\right)}^{k}\;.
\ee
This has exactly the same structure as a term in the OPE expansion, Eq.(26), and one sees that
the branch point exponent $\gamma$ of the $IR$ renormalon is related to the anomalous dimension
of the operator, with ${\delta}_{k}=1-{\gamma}_{k}$.
 The idea is that the coefficient, ${\cal{C}}_{k}$,
 in particular the constant $K$, is ambiguous in the OPE because of non-logarithmic UV
divergences \cite{r25,r27}. This ambiguity can be compensated by the IR renormalon ambiguity in the
PT Borel integral, and so regulating the Borel integral, using for instance a principal value (PV)
prescription, induces a particular definition of the coefficient functions in the OPE, and the
PT and OPE components are then separately well-defined. That this scenario works in detail
can be confirmed in toy models such as the non-linear $O(N)$ ${\sigma}$-model \cite{r25,r28}. For the QED
case the ambiguity corresponds to a ${Q}^{2}/{\tilde{\Lambda}}^{2}$ effect. So that all-orders
QED perturbation theory is only defined if there are in addition power corrections in ${Q}^{2}$.
Such effects are only important if ${Q}^{2}{\sim}{\tilde{\Lambda}}^{2}$, here ${\tilde{\Lambda}}$ corresponds
to the Landau ghost in QED, ${\tilde{\Lambda}}^{2}\sim{10}^{560}{m}^{2}$, with $m$ the fermion mass.
Thus in QED such power corrections can have no phenomenological consequences and can be completely
ignored.\\

Our exact information about the Borel transform, $B[{\cal{D}}](z)$, for the QCD Adler $D$ function
is restricted to the large-$N_f$ result of Eq.(29). In QCD we expect large-order behaviour in
perturbation theory of the form ${d}_{n}{\approx}K{n}^{\gamma}{(b/2)}^{n}{n!}$, involving the
QCD beta-function coefficient $b=(33-2{N}_{f})/6$. Motivated by the structure of renormalon
singularities in QCD one can then convert the ${N}_{f}$ expansion into the so-called $b$-expansion \cite{r29,r29a,r30,r31},
by substituting ${N}_{f}=(33/2-3b)$ to obtain,
\be
{d}_{n}={d}_{n}^{(n)}{b}^{n}+{d}_{n}^{(n-1)}{b}^{n-1}+\ldots+{d}_{n}^{(0)}\;.
\ee
The leading-$b$ term ${d}_{n}^{(L)}{\equiv}{d}_{n}^{(n)}{b}^{n}$ is then used to approximate $d_n$.
Since ${d}_{n}^{(L)}=
{(-3)}^{n}{d}_{n}^{[n]}{b}^{n}$, it is known to all-orders from the large-$N_f$ result.
This approach is sometimes also referred to as ``Naive Nonabelianization'' \cite{r29}. It
can be motivated by considering a QCD skeleton expansion \cite{r32}, and corresponds to
simply taking the first ``one-chain'' term in the expansion. It does not include the
multiple exchanges of renormalon chains needed to build the full asymptotic behaviour of
the perturbative coefficients, and there are no firm guarantees as to its accuracy. The
leading-$b$ result for the Borel transform of the Adler-$D$ function in the $V$-scheme
can then be obtained from Eq.(29).
\be
B[{\cal{D}}^{(L)}](z)={\sum_{j=1}^{\infty}}\frac{{A}_{0}(j)+z{A}_{1}(j)}{{(1+\frac{z}{{z}_{j}})}^{2}}+
\frac{{B}_{0}(2)}{(1-\frac{z}{z_2})}+{\sum_{j=3}^{\infty}}\frac{{B}_{0}(j)+z{B}_{1}(j)}{{(1-\frac{z}{z_j})}^{2}}\;,
\ee
so that one sees in the leading-$b$ limit a set of single and double pole renormalon singularities at the
expected positions. The residues of the ${UV}_{j}$ poles, ${A}_{0}(j)$ and ${A}_{1}(j)$, are given by \cite{r29a}
\be
{A}_{0}(j)=\frac{8}{3}\frac{{(-1)}^{j+1}(3{j}^{2}+6j+2)}{{j}^{2}{(j+1)}^{2}{(j+2)}^{2}}\;\;
{A}_{1}(j)=\frac{4}{3}\frac{b{(-1)}^{j+1}(2j+3)}{{j}^{2}{(j+1)}^{2}{(j+2)}^{2}}\;.
\ee
Because of the conformal symmetry of the vector correlator \cite{r33} the ${IR}_{j}$ residues, ${B}_{0}(j)$
and ${B}_{1}(j)$, are directly related to the ${UV}_{j}$ ones, with ${B}_{0}(j)=-{A}_{0}(-j)$ and
${B}_{1}(j)=-{A}_{1}(-j)$ for $j>2$. ${B}_{0}(1)={B}_{1}(1)={B}_{1}(2)=0$, and ${B}_{0}(2)=1$. Notice
the absence of an ${IR}_{1}$ renormalon singularity. This is consistent with the correspondence  between
OPE terms and IR renormalon ambiguities noted above, since there is no relevant operator of dimension 2
in the OPE. The singularity nearest the origin is then the ${UV}_{1}$ singularity at $z=-2/b$, which
generates the leading asymptotic behaviour,
\be
{d}_{n}^{(L)}{\approx}\frac{(12n+22)}{27}{n!}{\left(\frac{-b}{2}\right)}^{n}\;.
\ee

We shall now consider the correction, ${\cal{R}}(s)$, to the parton model result for ${R}_{{e}^{+}{e}^{-}}$.
This may be split into a perturbative component ${\cal{R}}_{PT}(s)$, and an OPE component ${\cal{R}}_{NP}(s)$,
analogous to Eqs.(23),(24). Inserting the Borel representation for ${\cal{D}}_{PT}$ of Eq.(30) into
the dispersion relation of Eq.(7) one finds the representation
\be
{\cal{R}}_{PT}(s)=\frac{1}{2\pi{i}}{\int_{-s-i\epsilon}^{-s+i\epsilon}}\frac{dt}{t}{\int_{0}^{\infty}}
{dz}{e}^{{-z}/{a}(t)}B[{\cal{D}}](z)\;.
\ee
It will be convenient to consider the all-orders perturbative result in leading-$b$ approximation
to start with, in which case the coupling $a(t)$ will have its one-loop form, $a(t)=2/(b{\ln}(t/{\tilde{\Lambda}}_{V}^{2}))$,
where we assume the $V$-scheme. In this case the $t$ integration is trivial and one finds,
\be
{\cal{R}}^{(L)}_{PT}(s)={\int_{0}^{\infty}}{dz}{e}^{{-z}/a(s)}\frac{{\sin}({\pi}bz/2)}{{\pi}bz/2}B[{\cal{D}}^{(L)}](z)\;,
\ee
where $B[{\cal{D}}^{(L)}](z)$ (in the $V$-scheme) is given by Eq.(35). It is now possible to explicitly
evaluate ${\cal{R}}^{(L)}_{PT}(s)$ in terms of generalised exponential integral functions $Ei(n,w)$,
defined for $Rew>0$ by
\be
Ei(n,w)=\int_{1}^{\infty}{dt}\frac{{e}^{-wt}}{t^n}\;.
\ee
One also needs the integral
\be
\int_{0}^{\infty}{dz}{e}^{-z/a}\frac{{\sin}(\pi{b}z/2)}{z}={\arctan}\left(\frac{{\pi}ba}{2}\right)\;.
\ee
Writing the `$\sin$' as a sum of complex exponentials and using partial fractions one can then evaluate
the contribution to ${\cal{R}}^{(L)}_{PT}(s)$ coming from the UV renormalon singularities, i.e. from
the terms involving ${A}_{0}(j)$ and ${A}_{1}(j)$ in Eq.(35) \cite{r29a}
\ba
{\cal{R}}_{PT}^{(L)}(s){|}_{UV}&=&\frac{2}{{\pi}b}\left(\frac{8{\zeta}_{2}}{3}-\frac{11}{3}\right)
{\arctan}\left(\frac{{\pi}ba(s)}{2}\right)
\nonumber \\
&+&\frac{2}{{\pi}b}\sum_{j=1}^{j=\infty}\left({A}_{0}(j){\phi}_{+}(1,j)+({A}_{0}(j)-{A}_{1}(j){z}_{j})
{\phi}_{+}(2,j)\right)\;,
\ea
where ${\zeta}_{2}={\pi}^{2}/6$ is the Riemann zeta-function, and we have defined
\be
{\phi}_{+}(p,q){\equiv}{e}^{{z}_{q}/a(s)}{(-1)}^{q}Im[Ei(p,(1/a(s))+i{\pi}b{z}_{q}/2)]\;.
\ee
To evaluate the remaining contribution involving the IR renormalon singularities we need to
regulate the integral to deal with the singularities on the integration contour. For simplicity
we could choose to take a principal value prescription. We need to continue the $Ei(n,w)$ defined
for $Rew>0$ by Eq.(40), to $Rew<0$. With the standard continuation one arrives at a function
analytic everywhere in the cut complex $w$-plane, except at $w=0$; with a branch cut running
along the negative real axis. Explicitly \cite{r34}
\be
Ei(n,w)=\frac{{(-w)}^{n-1}}{(n-1)!}\left[-{\ln}w-{\gamma}_{E}+\sum_{m=1}^{n-1}\frac{1}{m}\right]
-\sum_{\scriptstyle{m}={0}\atop\scriptstyle{m}\neq{n-1}}\frac{{(-w)}^{m}}{(m-n+1)m!}\;,
\ee
with ${\gamma}_{E}=0.5722{\ldots}$ Euler's constant. The $\ln{w}$ contributes the branch
cut along the negative real $w$-axis. To obtain the principal value of the Borel integral
one needs to compensate for the discontinuity across the branch cut, and make the replacement
${Ei}(n,w)\rightarrow{Ei}(n,w)+i{\pi}sign(Imw)$. This leads one to introduce, analogous to Eq.(43),
\ba
{\phi}_{-}(p,q)&{\equiv}&{e}^{-{z}_{q}/{a}(s)}{(-1)}^{q}Im[Ei(p,(-1/a(s))-i{\pi}b{z}_{q}/2)]
\nonumber \\
&-&\frac{{e}^{-{z}_{q}/{a}(s)}{(-1)}^{q}{z}_{q}^{p-1}}{(p-1)!}{\pi}Re[{((1/{a}(s))+i{\pi}b/2)}^{p-1}]\,.
\ea
The principal value of the IR renormalon contribution is then given by \cite{r29a}
\ba
{\cal{R}}^{(L)}_{PT}(s){|}_{IR}&=&\frac{2}{\pi{b}}\left(\frac{14}{3}-\frac{8{\zeta}_{2}}{3}\right)
{\arctan}\left(\frac{\pi{b}{a}(s)}{2}\right)+\frac{2{B}_{0}(2)}{{\pi}b}{\phi}_{-}(1,2)
\nonumber \\
&+&\frac{2}{{\pi}b}\sum_{j=3}^{\infty}\left({B}_{0}(j){\phi}_{-}(1,j)+({B}_{0}(j)+{B}_{1}(j){z}_{j})
{\phi}_{-}(2,j)\right)\;.
\ea
The perturbative component is then the sum of the UV and (regulated) IR contributions,
\ba
{\cal{R}}^{(L)}_{PT}(s)&=&{\cal{R}}^{(L)}_{PT}(s){|}_{UV}+{\cal{R}}^{(L)}_{PT}(s){|}_{IR}
\nonumber \\
&=&\frac{2}{{\pi}b}{\arctan}\left(\frac{{\pi}b{a(s)}}{2}\right)+\frac{2}{{\pi}b}\sum_{j=1}^{\infty}\left({A_0}(j){\phi}_{+}(1,j)
+({A_0}(j)-{A_1}(j){z}_{j}){\phi}_{+}(2,j)\right)
\nonumber \\
&+&\frac{2{B_0}(2)}{{\pi}b}{\phi}_{-}(1,2)+\frac{2}{{\pi}b}\sum_{j=3}^{\infty}\left({B_0}(j){\phi}_{-}(1,j)
+({B_0}(j)+{B_1}(j){z}_{j}){\phi}_{-}(2,j)\right)\;.
\ea
Note that the ${\zeta}_{2}$ contributions cancel, and one obtains the ${\arctan}$ term, which is the leading
contribution, ${A}_{1}(s)$, in the CIPT/APT reformulation of fixed-order perturbation theory. The connection
between the Borel representation and the ${A}_{n}(s)$ will be further clarified later.\\

We now turn to the infrared behaviour of the regulated Borel integral. In the one-loop
(leading-$b$) case the $V$-scheme coupling $a(s)$ becomes infinite at $s={s}_{L}{\equiv}{\tilde{\Lambda}}_{V}^{2}$.
The ${e}^{-z/{a}(s)}$ term in the Borel integrand approaches unity at $s={s}_{L}$, but the trigonometric factor
${\sin}({\pi}bz/2)/({\pi}bz/2)$ ensures that the integral is defined at ${s}={s}_{L}$. For $s<{s}_{L}$,
however, ${a}(s)$ becomes negative, and the ${e}^{-z/{a}(s)}$ factor diverges at $z=\infty$, the Borel
transform in the $V$-scheme does not contain any exponential $z$-dependence to compensate,
 so the
Borel integral is not defined. We shall refer to this pathology of the Borel integral at ${s}={s}_{L}$
as the ``Landau divergence''. It is important to stress that the Landau divergence is to be carefully
distinguished from the Landau pole in the coupling. The Landau pole in the coupling depends on the
chosen renormalization scale. At one-loop choosing an ${\overline{MS}}$ scale ${\mu}^{2}=xs$, the
coupling $a(xs)$ has a Landau pole at $s={\tilde{\Lambda}}_{\overline{MS}}^{2}/x$, the Borel integral
of Eq.(39) can then be written in terms of this coupling as,
\be
{\cal{R}}^{(L)}_{PT}(s)=\int_{0}^{\infty}{dz}{e}^{-z/a(xs)}\frac{{\sin}({\pi}bz/2)}{{\pi}bz/2}
{[x{e}^{5/3}]}^{bz/2}B[{\cal{D}}^{(L)}](z)\;.
\ee
In a general scheme the Borel transform picks up the extra factor ${[x{e}^{5/3}]}^{bz/2}$ multiplying
the $V$-scheme result. The Borel integrand is scheme ($x$) invariant. The extra factor has to be
taken into account when identifying where the integral breaks down, and one of course finds the
Landau divergence to be at the same $x$-independent energy, $s={s}_{L}={e}^{5/3}{\tilde{\Lambda}}_{\overline{MS}}^{2}=
{\tilde{\Lambda}}_{V}^{2}$. Thus the Borel representation of Eq.(38) for ${\cal{R}}^{(L)}_{PT}(s)$ only applies for
$s{\geq}{s}_{L}$.
 For $s<{s}_{L}$ the one-loop ($V$-scheme) coupling
$a(s)$ becomes negative. We can rewrite the perturbative expansion of ${\cal{R}}_{PT}(s)$ as an
expansion in $(-a(s))$,
\ba
{\cal{R}}_{PT}(s)&=&a(s)+{r_1}{a}^{2}(s)+{r_2}{a}^{3}(s)+\ldots+{r_n}{a}^{n+1}(s)+\ldots
\nonumber \\
&=&-[(-a(s))-{r_1}{(-a(s))}^{2}+{r_2}{(-a(s))}^{3}+{\ldots}+{(-1)}^{n}{r_n}{(-a(s))}^{n+1}+\ldots]\;.
\ea
The expansion in $(-a(s))$ follows from the modified Borel representation
\ba
{\cal{R}}_{PT}(s)&=&-{\int_{0}^{\infty}}{dz}{e}^{-z/(-a(s))}B[{\cal{R}}](-z)
\nonumber \\
&=&\int_{0}^{-\infty}{dz}{e}^{z/(-a(s))}B[{\cal{R}}](z)\;.
\ea
This modified form of Borel representation will be valid when $Re(a(s))<0$, and involves
an integration contour along the negative real axis. Thus, it is now the {\it ultraviolet}
renormalons ${UV}_{k}$ which render the Borel integral ambiguous. The ambiguity in taking the contour
around these singularities (analogous to Eq.(33)) now involves ${(s/{\tilde{\Lambda}}^{2})}^{k}$.
Of course, it is now unclear how these ambiguities can cancel against the corresponding
OPE ambiguities. The key point is that since only the sum of the PT and OPE components is
well-defined, the Landau divergence of the Borel integral at $s={s}_{L}$, must be accompanied
by a corresponding breakdown in the validity of the OPE as an expansion in powers
of ${({\tilde{\Lambda}}^{2}/s)}$, at the same energy. The idea is illustrated by the
following toy example, where the OPE is an alternating geometric progression,
\ba
{\cal{R}}_{NP}(s)&=&{\left(\frac{{\tilde{\Lambda}}^{2}}{s}\right)}-{\left(\frac{{\tilde{\Lambda}}^{2}}{s}\right)}^{2}
+{\left(\frac{{\tilde{\Lambda}}^{2}}{s}\right)}^{3}-\ldots
\nonumber \\
&=&\frac{\frac{{\tilde{\Lambda}}^{2}}{s}}{1+\frac{{\tilde{\Lambda}}^{2}}{s}}=\frac{1}
{1+\frac{{s}}{\tilde{\Lambda}^{2}}}
\nonumber \\
&=&1-
{\left(\frac{s} {{\tilde{\Lambda}}^{2}}\right)}+{\left(\frac{s}{{\tilde{\Lambda}}^{2}}\right)}^{2}
-{\left(\frac{s}{{\tilde{\Lambda}}^{2}}\right)}^{3}-\ldots\;.
\ea
At any value of $s$, ${\cal{R}}_{NP}(s)$ is given by the equivalent functions in the middle line.
For $s>{\tilde{\Lambda}}^{2}$ these have a valid expansion in powers of ${\tilde{\Lambda}}^{2}/s$,
the standard OPE, given in the top line. For $s<{\tilde{\Lambda}}^{2}$ the standard OPE breaks down,
but there is a valid expansion in powers of ${s}/{\tilde{\Lambda}}^{2}$ given in the bottom line.
Thus for $s<{s}_{L}$ the OPE should be resummed and recast in the form,
\be
{\cal{R}}_{NP}(s)=\sum_{n}{\tilde{\cal{C}}}_{n}{\left(\frac{s}{{\tilde{\Lambda}}^{2}}\right)}^{n}\;.
\ee
It is crucial to note that this reorganised OPE can contain a ${\tilde{\cal{C}}}_{0}$ term which
is independent of $s$, as indeed is the case in the toy example of Eq.(51). Of course, an analogous
${\cal{C}}_{0}$ term in the standard OPE in Eq.(26) is clearly excluded since it would violate
Asymptotic Freedom, and all the terms in the regular OPE are perturbatively invisible. As a result
${\cal{R}}_{NP}(s)$ can have a non-vanishing infrared limit, and both components can contribute
to the infrared freezing behaviour. It should be no surprise that perturbation theory by itself
cannot determine the infrared behaviour of observables, but the existence of a well-defined
perturbative component which, as we shall claim, can be computed at all values of the energy
using a reorganised APT version of fixed-order perturbation theory, is a noteworthy feature.
The remaining terms present in this modified OPE should then be in one-to-one correspondence with the
${UV}_{n}$ renormalon singularities in the Borel transform of the PT component, and the PT
renormalon ambiguities can cancel against corresponding OPE ones, and again each component
separately be well-defined. The modified coefficients ${\tilde{\cal{C}}}_{n}$ will have
a form analogous to Eq.(27),
\be
{\tilde{\cal{C}}}_{n}=K{a}^{\tilde{\delta}_{n}}({\mu}^{2})[1+O(a)]\;.
\ee
The anomalous dimension is that of an operator which can be identified using
the technique of Parisi \cite{r23}. The anomalous dimension corresponding to $\tilde{\cal{C}}_{1}$
for the Adler $D$ function has been computed \cite{r35}. The ambiguity for the modified
Borel representation of Eq.(50), taking ${UV}_{k}$ to be a branch point singularity
${(1-z/{z}_{k})}^{{\tilde{\gamma}}_{k}}$, is
\be
{\Delta}{R}_{PT}{\approx}K{a}^{1-{\tilde{\gamma}}_{k}}{\left(\frac{s}{{\tilde{\Lambda}}^{2}}\right)}^{k}\;.
\ee
Comparing with Eq.(53) one finds ${\tilde{\delta}}_{k}=1-{\tilde{\gamma}}_{k}$.
 The modified Borel representation for ${\cal{R}}_{PT}^{(L)}$
valid for $s<{s_L}$ will be,
\be
{\cal{R}}_{PT}^{(L)}(s)=-\int_{0}^{\infty}{dz}{e}^{-z/(-a(s))}B[{\cal{R}}^{(L)}](-z)\;.
\ee
This may again be written explicitly in terms of ${Ei}(n,w)$ functions. One simply
needs to change ${a(s)}\rightarrow{-a(s)}$, ${z_j}\rightarrow{-{z}_{j}}$, and
${A_1}(j)\rightarrow{-{A_1}(j)}$, ${B_1}(j)\rightarrow{-{B_1}(j)}$ in Eq.(47). One finds
that the result of Eq.(47) is invariant under these changes, apart from the
additional terms which we added to the ${Ei}(n,w)$ in continuing from $Rew>0$ to
$Rew<0$, in order to obtain the principal value. In fact the PV Borel integral
is not continuous at $s={s}_{L}$. Continuity is obtained if rather than the
principal value we use the standard continuation of the ${Ei}(n,w)$ defined by
Eq.(44). That is we redefine
\be
{\phi}_{-}(p,q){\equiv}{e}^{-{z}_{q}/{a}(s)}{(-1)}^{q}Im[Ei(p,(-1/a(s))-i{\pi}b{z}_{q}/2)]\;.
\ee
This simply corresponds to a different regulation of singularities. We then see that
Eq.(47) for ${\cal{R}}^{(L)}_{PT}(s)$ is a function of $a(s)$ which is well-defined at
all energies, and freezes to $2/b$ in the infra-red. We note that the branch of the
$\arctan$ changes at $s={s}_{L}$, so that its value smoothly changes from zero
at $s=\infty$ to ${\pi}$ at $s=0$. The reformulated OPE of Eq.(52) together with the
perturbative component determines the infrared freezing behaviour, and in
the ultraviolet the perturbative component dominates. The key point
is that both components can be described by functions of $s$ which are well-defined at
all energies. The apparent Landau divergence simply reflects the fact that the
Borel integral and OPE series, which are used to describe the PT and NP components, each have a limited
range of validity in $s$.
 The connection with the CIPT/APT rearrangement of
fixed-order perturbation theory is now clear. It is obtained by keeping the ${\sin}(\pi{b}z/2)/({\pi}bz/2)$
term in the Borel transform intact, and expanding the remainder in powers of $z$. Ordinary fixed-order
perturbation theory, of course, corresponds to expanding the whole Borel transform in powers
of $z$. The retention of the oscillatory $\sin$ factor in the Borel transform ensures that the
reformulated perturbation theory remains defined at all energies.
 One then finds that for
$s{\geq}{s_L}$,
\be
{A}_{n}(s)=\int_{0}^{\infty}{dz}{e}^{-z/a(s)}\frac{{\sin}({\pi}bz/2)}{{\pi}bz/2}\frac{{z}^{n-1}}{(n-1)!}\;,
\ee
where the one-loop ${A}_{n}(s)$ are given by Eqs.(12). Similarly for $s{\leq}{s}_{L}$ one finds
\be
{A}_{n}(s)=\int_{0}^{-\infty}{dz}{e}^{z/(-a(s))}\frac{{\sin}({\pi}bz/2)}{{\pi}bz/2}\frac{{z}^{n-1}}{(n-1)!}\;.
\ee
Thus the CIPT/APT fixed-order result should be an asymptotic approximation to the Borel integral
at both large and small values of $s$. In Fig.4 we compare the all-orders leading-$b$ result
for ${\cal{R}}_{PT}^{(L)}(s)$ given by Eq.(47), with the NNLO CIPT/APT prediction,
\be
{\cal{R}}^{(L)}_{APT}(s)={A}_{1}(s)+{d}_{1}^{(L)}{A}_{2}(s)+{d}_{2}^{(L)}{A}_{3}(s)\;.
\ee
The one-loop ${A}_{n}(s)$ are given by Eqs.(12) and as in Eq.(47) the $V$-scheme is assumed. We
assume ${N_f}=2$ quark flavours.
One sees that there is good agreement at all values of ${s}/{\tilde{\Lambda}}_{V}^{2}$.\\

We now turn to the full QCD result beyond the one-loop approximation, and as in Section 2
it will be sufficient to consider the two-loop result since one can always use an 't Hooft
scheme. Consider the Borel representation for ${\cal{R}}_{PT}(s)$ of Eq.(38). We shall assume that,
as in the leading-$b$ approximation,
the Borel transform $B[{\cal{D}}](z)$ in the $V$-scheme does not contain any exponential
dependence on $z$, but is simply a combination of branch point singularities. It is then clear
that the Landau divergence occurs when the factor ${e}^{-z/{a}(-s)}$ becomes a diverging
exponential, that is when $Re(1/a(-s))<0$. Thus the critical energy ${s}_{L}$ is given
by the condition $Re(1/a(-s))=0$. At one-loop level one has
\be
\frac{1}{a(-s)}=\frac{b}{2}{\ln}\left(\frac{s}{{\tilde{\Lambda}}_{V}^{2}}\right)+\frac{i{\pi}b}{2}\;,
\ee
and so the condition yields $s={s}_{L}={\tilde{\Lambda}}_{V}^{2}$, as we found before.
At the two-loop level the situation is slightly different. Integrating the two-loop
beta-function in Eq.(14) now gives,
\be
\frac{1}{a(-s)}+c{\ln}\left[\frac{ca(-s)}{1+ca(-s)}\right]=\frac{b}{2}{\ln}\left(\frac{s}{{\tilde{\Lambda}}_{V}^{2}}
\right)+\frac{i{\pi}b}{2}\;.
\ee
The vanishing of $Re(1/a(-s))$ then corresponds to the solution of the transcendental equation
\be
Re\left\{ c{\ln}
\left[\frac{ca(-s)}{1+ca(-s)}\right]
\right\}=\frac{b}{2}{\ln}\left(\frac{s}{{\tilde{\Lambda}}_{V}^{2}}\right)
\;.
\ee
Assuming ${N}_{f}=2$ flavours one finds $s={s}_{L}=0.4574{\tilde{\Lambda}}_{V}^{2}$. Since the Borel integral
is scheme-invariant so must the value of $s_L$ be, in particular the breakdown of the
Borel representation would occur in any scheme, not just an 't Hooft one. We can perform the
$t$-integration in Eq.(38) in closed form, and arrive at the two-loop Borel representation
\be
{\cal{R}}_{PT}(s)=\frac{-2}{{\pi}b}\int_{0}^{\infty}{dz}Im\left[\frac{{e}^{-z/a(-s+i\epsilon)}}{z}-c{e}^{zc}Ei\left(1,
zc+\frac{z}{a(-s+i\epsilon)}\right)\right]B[{\cal{D}}](z)\;.
\ee
The factor in the square bracket plays the role of the ${e}^{-z/a(s)}{\sin}({\pi}bz/2)/({\pi}bz/2)$
factor in the one-loop case. It provides an oscillatory factor so that at $s={s}_{L}$ the
Borel representation remains defined. For ${s}<{s}_{L}$ one must switch to a modified Borel
representation as in Eq.(50), writing
\be
{\cal{R}}_{PT}(s)=-\frac{1}{2\pi{i}}{\int_{-s-i\epsilon}^{-s+i\epsilon}}\frac{dt}{t}{\int_{0}^{\infty}
{dz}{e}^{{-z}/(-{a}(t))}}B[{\cal{D}}](-z)\;.
\ee
Which, performing the $t$-integration gives
\ba
{\cal{R}}_{PT}(s)&=&\frac{2}{{\pi}b}\int_{0}^{\infty}{dz}Im\left[-\frac{{e}^{-z/(-a(-s+i\epsilon))}}{z} \right.
  \nonumber \\  &-& \left. c{e}^{-zc}Ei\left(1,
-zc+\frac{z}{(-a(-s+i\epsilon))}\right)\right]B[{\cal{D}}](-z)\;.
\ea
Unfortunately we cannot write down a function analogous to Eq.(47) which gives
${\cal{R}}_{PT}(s)$ at all energies, because we do not know
$B[{\cal{D}}](z)$ exactly. The two-loop situation, however, is the same as
that at one-loop. The regulated representation of Eq.(63) applies for $s{\geq}{s}_{L}$
, with the corresponding standard OPE. Below $s={s}_{L}$ one needs the modified
representation of Eq.(65) together with the resummed OPE recast in the form of
Eq.(52). The perturbative component ${\cal{R}}_{PT}(s)$ then freezes to $2/b$
in the infra-red,
we can see this if we split $B[{\cal{D}}](-z)$ into $(1
+(B[{\cal{D}}](-z)-1))$. The part of the integrand proportional to
$B[{\cal{D}}](-z)-1$ vanishes for all $z$ from $0 \to \infty$ in the infra-red limit. The
remaining term integrates to give us $A_{1}(s)$ which freezes to $2/b$ as $s
\to 0$.
The non-perturbative component
${\cal{R}}_{NP}(s)$ given by the reformulated OPE together with the perturbative component
determine the infrared freezing behaviour.
There is again a direct connection with the CIPT/APT reformulation of fixed-order
perturbation theory. Using integration by parts one can show that
that for $s{\geq}{s}_{L}$
\be
{A}_{n}(s)=\frac{-2}{{\pi}b}\int_{0}^{\infty}{dz}Im\left[\frac{{e}^{-z/a(-s+i\epsilon)}}{z}-c{e}^{zc}Ei\left(1,
zc+\frac{z}{a(-s+i\epsilon)}\right)\right]\frac{{z}^{n-1}}{(n-1)!}\;,
\ee
where the ${A}_{n}(s)$ correspond to the two-loop results in Eqs.(20,21). Once again CIPT/APT
corresponds to keeping the oscillatory function in the Borel transform intact, and expanding the
remainder in powers of $z$.
 Similarly for
$s{\leq}{s}_{L}$ one has,
\ba
{A}_{n}(s)&=&\frac{2}{{\pi}b}\int_{0}^{\infty}{dz}Im\left[-\frac{{e}^{-z/(-a(-s+i\epsilon))}}{z} \right.
  \nonumber \\
&-& \left. c{e}^{-zc}Ei\left(1,
-zc+\frac{z}{(-a(-s+i\epsilon))}\right)\right]\frac{{(-z)}^{n-1}}{(n-1)!}\;.
\ea

\begin{figure}
\begin{center}
 \psfrag{x}[t][bl]{$s/\tilde{\Lambda}_{V}^2$}%
 \psfrag{y}[tr][br]{$\delta{\cal{R}}(s)$}

\epsfig{file=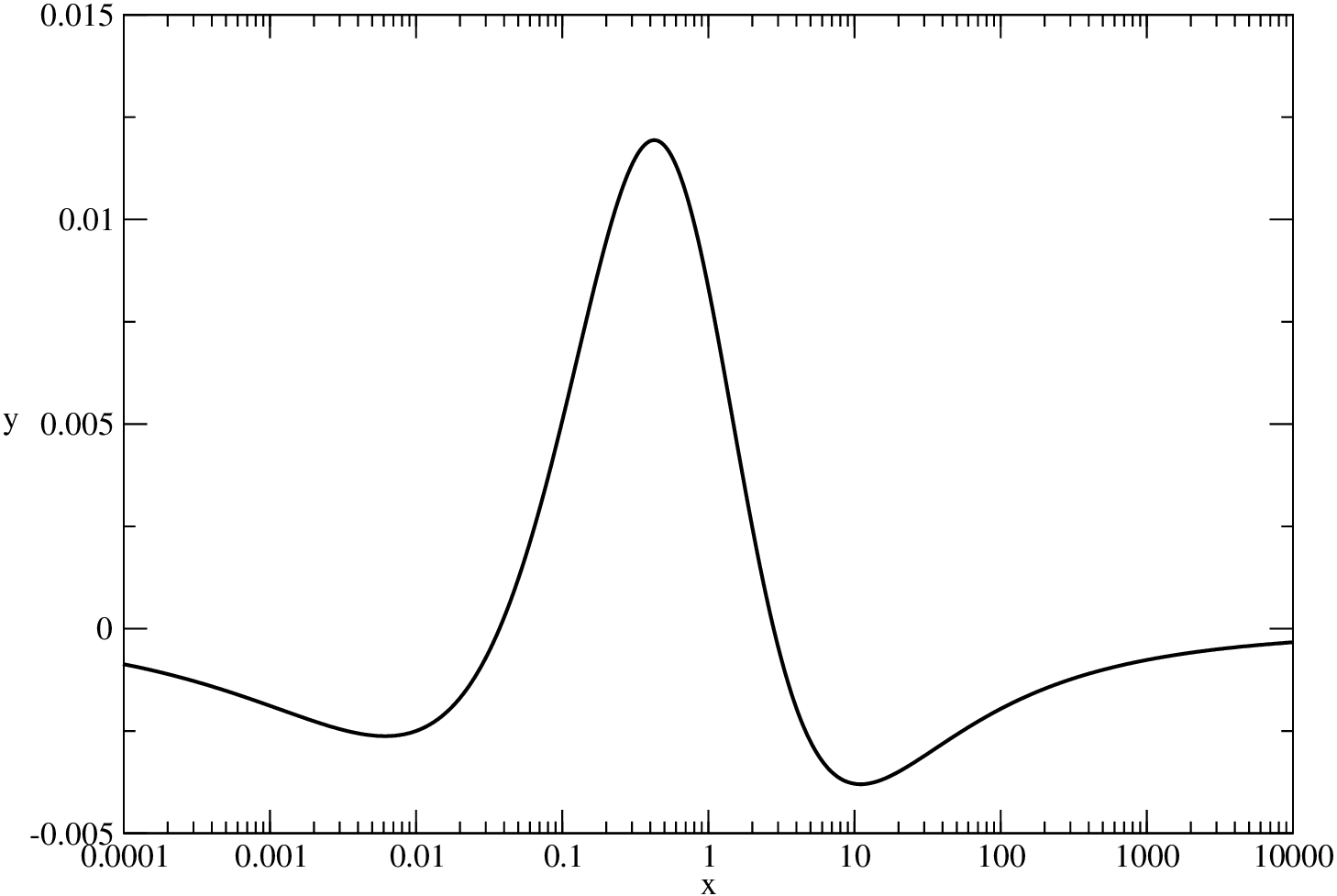,height=7.0cm,width=10.5cm}
\caption{$\delta{\cal{R}}(s) = {\cal{R}}^{(L)}_{PT}(s)-{\cal{R}}^{(L)}_{APT}(s)$,
at the one loop level for 2
flavours of quark.}
\end{center}
\end{figure}

Thus, as in the one-loop case, the CIPT/APT reformulation of fixed-order perturbation
theory will be asymptotic to the Borel representations at small and large energies.
We would like, as in Fig.4 for the one-loop case, to compare how well the fixed-order
CIPT/APT perturbation theory corresponds with the all-orders Borel representation. We are
necessarily restricted to using the leading-$b$ approximation since this is the extent
of the exact all-orders information at our disposal. One possibility is to simply use
the leading-$b$ result for the Borel transform, ${B}[{\cal{D}}^{(L)}](z)$, in the
two-loop Borel representation of Eq.(63). The difficulty though is that with $a(-s)$
the two-loop coupling, the Borel integral is now scheme-dependent, since $B[{\cal{D}}^{(L)}](z)$
has a scale dependence which exactly compensates that of the {\it one}-loop coupling. Using
a renormalization scale ${\mu}^{2}=xs$ our result for ${\cal{R}}_{PT}(s)$ has an
unphysical $x$-dependence.
 This difficulty is exacerbated if we attempt to
match the result to the exactly known perturbative coefficients $d_1$ and $d_2$,
which we could do by adding an additional contribution $({d_1}-{d}_{1}^{(L)})z+
({d_2}-{d}_{2}^{(L)})({z}^{2}/2)$ to the Borel transform. Thus, as has been argued elsewhere,
such matching of leading-$b$ results to exact higher-order results yields completely {\it ad hoc} predictions,
which may be varied at will by changing the renormalisation scale \cite{r36,r37}. The resolution of
this difficulty follows if one accepts that the standard RG-improvement of fixed-order
perturbation theory is incomplete, in that only a subset of RG-predictable UV logarithms involving
the energy scale $s$ are resummed. Performing a complete resummation of these logs together
with the accompanying logs involving the renormalisation scale, yields a scale-independent
result. This Complete Renormalisation Group Improvement (CORGI) approach \cite{r39} applied to
${\cal{D}}(s)$ corresponds to use of a renormalisation scale ${\mu}^{2}={e}^{-2d/b}s$, where
$d$ denotes the NLO perturbative correction $d_1$ in Eq.(23), in the ${\overline{MS}}$ scheme
with ${\mu}^{2}=s$. In the CORGI scheme we have the perturbation series,
\be
{{\cal{D}}(t)}={a}_{0}(t)+{X}_{2}{a}_{0}^{3}(t)+{X}_{3}{a}_{0}^{4}(t)+\ldots+{X}_{n}{a}_{0}^{n+1}+\ldots\;,
\ee
where ${a}_{0}(t)$ is given by Eq.(15) with $z=(-1/e){(\sqrt{t}/{\Lambda}_{D})}^{-b/c}$, where
${\Lambda}_{D}{\equiv}{e}^{d/b}{\tilde{\Lambda}}_{\overline{MS}}$, and $X_n$ are the CORGI
invariants, and only $X_2$ is known.
 We can then attempt to
perform the leading-$b$ CORGI resummation,
\be
{\cal{D}}^{(L)}_{CORGI}(t)={a}_{0}(t)+{X}_{2}{a}_{0}^{3}(t)+\sum_{n>2}{X}_{n}^{(L)}{a}_{0}^{n+1}(t)+\ldots\;,
\ee
so that the exactly known NNLO $X_2$ coefficient is included, with the remaining unknown coefficients
approximated by ${X}_{3}^{(L)}$, ${X}_{4}^{(L)},\ldots$, the leading-$b$ approximations. We stress that
${a}_{0}(t)$ denotes the full CORGI coupling defined in Eq.(15).
 One can define this formal sum using the Borel representation
of ${\cal{D}}$ in Eq.(30), with the result for $B[{\cal{D}}^{(L)}]$ in Eq.(35). The integral can
be expressed in closed form in terms of the Exponential Integral functions $Ei(n,w)$ of Eq.(40),
with the result \cite{r7}
\ba
{\cal{D}}^{(L)}(1/a(t))&=&\sum_{j=1}^{\infty}{z}_{j}{\{}-{e}^{{z}_{j}/a(t)}Ei(1,{z}_{j}/a(t))[
({z}_{j}/a(t))({A}_{0}(j)-{z}_{j}{A}_{1}(j))-{z}_{j}{A}_{1}(j)]
\nonumber \\
&+&({A}_{0}(j)-{z}_{j}{A}_{1}(j)){\}}-{e}^{-{z}_{j}/a(t)}{B}_{0}(2)Ei(1,-{z}_{j}/a(t))
\nonumber \\
&+&\sum_{j=3}^{\infty}{\{}-{e}^{-{z}_{j}/a(t)}Ei(1,-{z}_{j}/a(t))[({z}_{j}/a(t))({B}_{0}(j)+
{z}_{j}{B}_{1}(j))]
\nonumber \\
&-&({B}_{0}(j)+{z}_{j}{B}_{1}(j)){\}}\;.
\ea
To define the infra-red renormalon contribution we have assumed the standard continuation
of $Ei(n,w)$ from $Rew>0$ to $Rew<0$, defined by Eq.(44). In \cite{r7} a principal
value was assumed, which corresponds to adding $-i{\pi}sign(Im({z}_{j}/a(t))$ to
the $Ei(1,-{z}_{j}/a(t))$ term. As we found for ${\cal{R}}_{PT}^{(L)}(s)$ the
principal value is not continuous at $s={s}_{L}$, whereas the standard continuation
is. The formal resummation in Eq.(69) then corresponds to \cite{r7},
\be
{\cal{D}}^{(L)}_{CORGI}(t)={\cal{D}}^{(L)}\left(\frac{1}{{a}_{0}(t)}+{d}_{1}^{(L)}(V)\right)\;
+({X}_{2}-{X}_{2}^{(L)}){a}_{0}^{3}(t)\;,
\ee
once again ${a}_{0}(t)$ is the full CORGI coupling, and ${d}_{1}^{(L)}(V)$ denotes the NLO
leading-$b$ correction in the $V$-scheme. Inserting ${\cal{D}}_{CORGI}(t)$ inside the
dispersion relation of Eq.(7) one can then define,
\be
{\cal{R}}^{(L)}_{CORGI}(s)=\frac{1}{2{\pi}i}\int_{-s-i{\epsilon}}^{-s+i{\epsilon}}{dt}
\frac{{\cal{D}}_{CORGI}^{(L)}(t)}{t}\;.
\ee
This can be evaluated numerically,
if we have ${\cal{R}}^{(L)}_{CORGI}(s_1)$ then we can obtain
\be
{\cal{R}}^{(L)}_{CORGI}(s_2) = {\cal{R}}^{(L)}_{CORGI}(s_1) +
\frac{1}{2{\pi}i}\left(\int_{-s_2-i\epsilon}^{-s_1-i\epsilon}dt\frac{{\cal{D}}^{(L)}_{CORGI}(t)}{t}
  + \int_{-s_1+i\epsilon}^{-s_2+i\epsilon}dt\frac{{\cal{D}}^{(L)}_{CORGI}(t)}{t}\right)\;.
\ee
If we set $s_1$ to be large enough we can evaluate
${\cal{R}}^{(L)}_{CORGI}(s_1)$ using the circular contour in the $s$-
plane, as in Eq.(8). Combining this circular integral with the integrals above and below
the real negative axis we arrive at ${\cal{R}}^{(L)}_{CORGI}(s_2)$ where
$s_2$ can be as far into the infrared as we want.
The all-orders CORGI result can be compared with the NNLO CIPT/APT CORGI result,
\be
{\cal{R}}_{APT}(s)={A}_{1}(s)+{X}_{2}{A}_{3}(s)\;.
\ee
Here the ${A}_{n}(s)$ are the two-loop results of Eqs.(20,21), with $A(s)=(-1/e){(\sqrt{s}/{\Lambda}_{D})}^{-b/c}$
in the CORGI scheme. Analogous to Fig.4 we plot in Fig.5 the comparison of the all-orders
and NNLO APT CORGI results, ${N}_{f}=2$ quark flavours are assumed. As in the one-loop
case there is extremely close agreement at all values of $s$. For the fits to
low-energy ${R}_{{e}^{+}{e}^{-}}(s)$ data to be presented in the next section,
therefore, we shall use the NNLO CORGI APT result.\\

\begin{figure}
\begin{center}
 \psfrag{x}[t][bl]{$s/\tilde{\Lambda}_{V}^2$}%
 \psfrag{y}[tr][br]{$\delta{\cal{R}}(s)$}

\epsfig{file=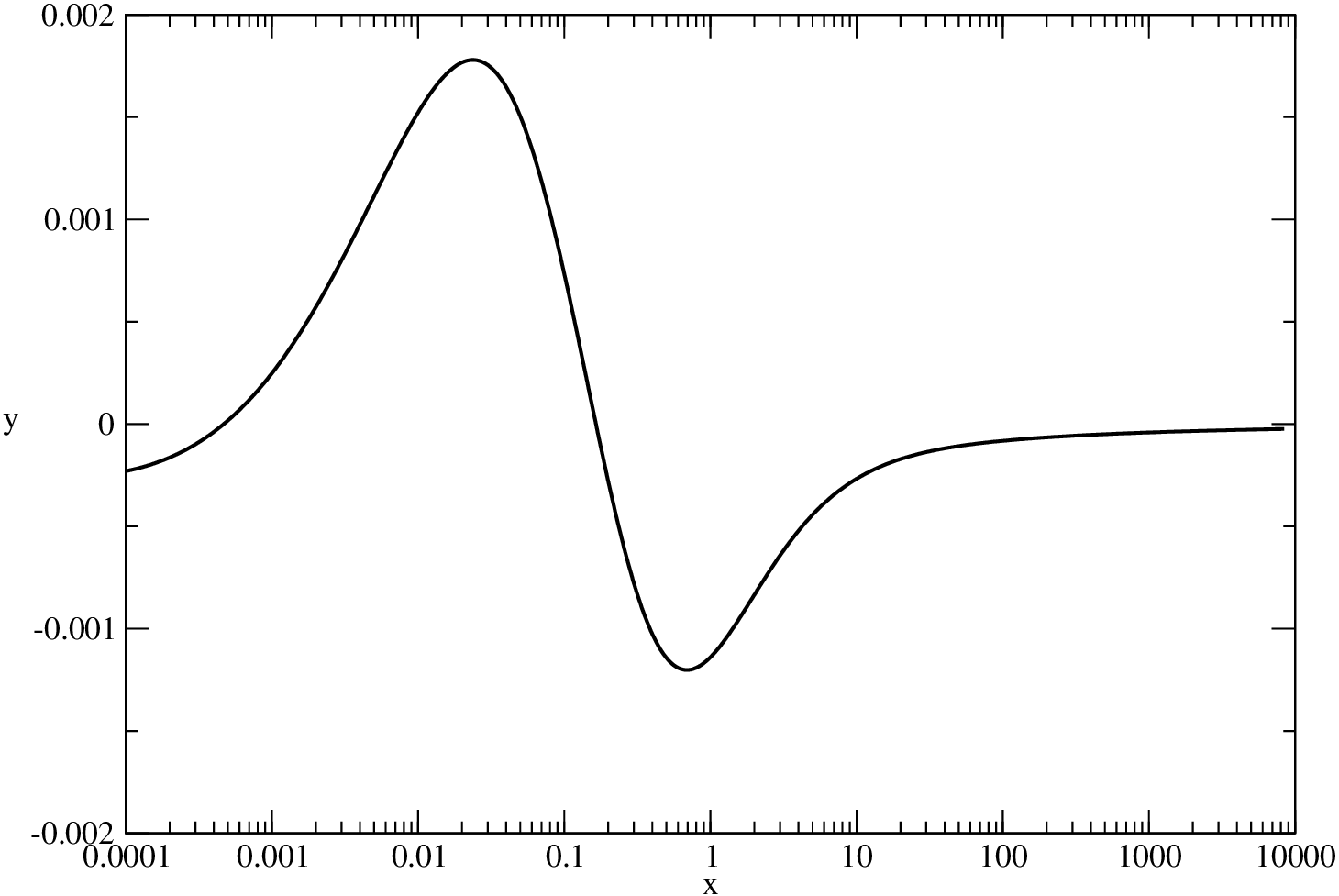,height=7.0cm,width=10.5cm}
\caption{$\delta{\cal{R}}(s) = {\cal{R}}^{(L)}_{CORGI}(s)-{\cal{R}}_{APT}(s)$,
 at the two loop level for 2
flavours of quark.}
\end{center}
\end{figure}

Before turning to phenomenological analysis in Section 4, we conclude this section
with a brief discussion of the situation for Euclidean observables. We can define
the Adler $D$ function in the Euclidean region by inverting the integral transform
corresponding to the dispersion relation of Eq.(7). That is we can write,
\be
{\cal{D}}(Q^2)={Q}^{2}\int_{0}^{\infty}\frac{ds}{{(s+{Q}^{2})}^{2}}{\cal{R}}(s)\;.
\ee
 One can certainly define a Euclidean version of APT by inserting
the Minkowskian ${A}_{n}(s)$ in the right-hand side of Eq.(75), and defining
\be
{A}^{(E)}_{n}({Q}^{2})={Q}^{2}\int_{0}^{\infty}\frac{ds}{{(s+{Q}^{2})}^{2}}{A}_{n}(s)\;.
\ee
The one-loop result would be \cite{apt1}
\be
{A}^{(E)}_{1}({Q}^{2})=\frac{2}{b}\left[\frac{1}{{\ln}({Q}^{2}/{\tilde{\Lambda}}^{2})}+
\frac{{\tilde{\Lambda}}^{2}}{{\tilde{\Lambda}}^{2}-{Q}^{2}}\right]\;.
\ee
This Euclidean APT coupling freezes in the infrared to $2/b$, but this behaviour is induced by the
second non-perturbative contribution, which cancels the forbidden tachyonic Landau pole
singularity present in the first perturbative term. There is now no direct connection, however,
between this Euclidean APT coupling and the Borel representation for ${D}_{PT}({Q}^{2})$ of Eq.(30).
Since there is now no oscillatory factor present in the Borel integral it is potentially divergent
at $s={s}_{L}$. We can explicitly exhibit this divergent behaviour working in leading-$b$ approximation.
The Borel integral can then be explicitly evaluated in terms of $Ei$ functions as we have seen in
Eq.(70). Using Eq.(44) for the $Ei$ function one then finds a divergent behaviour as $s\rightarrow{s}_{L}$
proportional to ${\ln}a$,
\be
{\cal{D}}_{PT}(s)\rightarrow [\sum_{j=1}^{\infty}({z}_{j}^{2}{A}_{1}(j)+{z}_{j}^{2}{B}_{1}(j))-{z}_{2}{B}_{0}(2)]
{\ln}a+\ldots\;,
\ee
where the ellipsis denotes terms finite as $s\rightarrow{s}_{L}$. However, remarkably, the factor
in the square bracket vanishes, and the result is finite at $s={s}_{L}$, provided that {\it all}
the renormalon singularities are included. The contribution of any individual renormalon is
divergent. The cancellation follows because of an exact relation between the residues of IR and
UV renormalons (Eq.(36)),
\be
{z}_{j}^{2}{A}_{1}(j)=-{z}_{j+3}^{2}{B}_{1}(j+3)\;.
\ee
This results in cancellations in the sum, leaving a residual term ${z}_{3}^{2}{B}_{1}(3)$ which
then cancels with the ${z}_{2}{B}_{0}(2)$ term. An analogous relation ${A}_{0}(j)=-{B}_{0}(j+2)$ has
been noted in \cite{r29a}. It seems that these relations are underwritten by the conformal symmetry
of the vacuum polarization function \cite{r33}, but further investigation is warranted. The above
finiteness at $s={s}_{L}$ means that one can obtain a ${\cal{D}}_{PT}({Q}^{2})$ component well-defined
in the infrared by changing to the modified form of Borel representation for $s<{s}_{L}$. One
finds that this component becomes negative before approaching the freezing limit ${\cal{D}}_{PT}(0)=0$.
Similar behaviour is found for the Gross-Llewellyn Smith and polarised and unpolarised Bjorken
structure function sum rules, whose complete renormalon structure is also known in leading-$b$
approximation \cite{r29a}. Phenomenological investigations are planned \cite{pb}. Comparable investigations
in the standard APT approach have been reported in \cite{apt2}.
 Unfortunately, nothing
is known about the full renormalon structure beyond leading-$b$ approximation. Such knowledge
would be tantamount to a full solution of the Schwinger-Dyson equations. Correspondingly no
analogue of the APT reorganisation of fixed-order perturbation theory asymptotic to
${\cal{D}}_{PT}$ is possible in the Euclidean case.

We finally note that in the case of ${R}_{{e}^{+}{e}^{-}}$ and ${D}$ it is possible to say
something about the separate infrared freezing behaviours of the PT and NP components. Arguments of
spontaneous chiral symmetry breaking in the limit of a large number of colours \cite{r33}
imply that $D(0)=0$, or equivalently ${\cal{D}}(0)=-1$. Furthermore
 according to
Ref.\cite{r44a} $\cal{R}$ and $\cal{D}$ should have the same infrared freezing limit.
This argument follows directly from Eq.(8) if the circular contour is shrunk to
zero. These exact results then suggest that ${\cal{D}}_{NP}(0)=-1$ to be consistent with
the leading-$b$ result ${\cal{D}}_{PT}(0)=0$ obtained above. For $\cal{R}$ one infers
that ${\cal{R}}_{NP}(0)=-1-(2/b)$ to be consistent with the ${\cal{R}}_{PT}(0)=2/b$ result.

\section*{4 Comparison of NNLO APT with low energy ${R}_{{e}^{+}{e}^{-}}$ data}

\begin{figure}
\begin{center}
\psfrag{x}[t][bl]{$s/\rm{GeV}^2$}
\psfrag{y}[tr][br]{$R_{e^+e^-}(s)$}

\epsfig{file=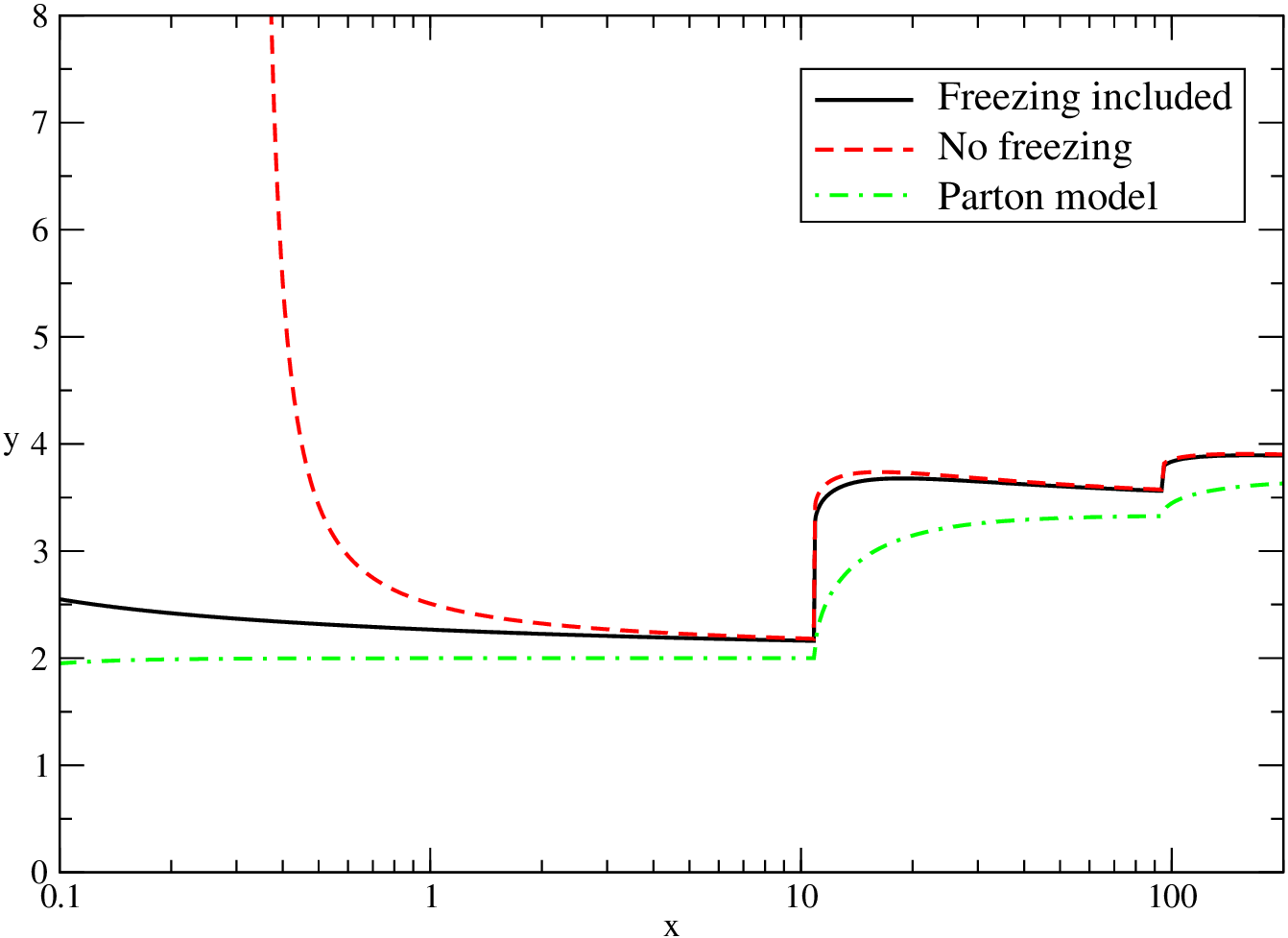,height=7.0cm,width=10.5cm}
\caption{Comparison of CORGI APT and the standard NNLO
 CORGI calculations of $R_{e^+e^-}(s)$ at
  low energies.}
\end{center}
\end{figure}
In this section we wish to compare the NNLO CORGI APT perturbative predictions
with low energy experimental data for ${R}_{{e}^{+}{e}^{-}}$. The discussion so
far has assumed massless quarks. To include quark masses we use the approximate
result \cite{r3,r40c}
\be
{R}_{{e}^{+}{e}^{-}}(s)=3\sum_{f}{Q}_{f}^{2}T({v}_{f})[1+g({v}_{f}){\cal{R}}(s)]\;,
\ee
with the sum over all active quark flavours, i.e. those with masses ${m}_{f}<\sqrt{s}/2$,
and where
\ba
{v}_{f}&=&{(1-4{m}_{f}^{2}/s)}^{\frac{1}{2}}\;,
\nonumber \\
T(v)&=&v(3-v^2)/2\;,
\nonumber \\
g(v)&=&\frac{4\pi}{3}\left[\frac{\pi}{2v}-\frac{3+v}{4}\left(\frac{\pi}{2}-\frac{3}{4\pi}\right)\right]\;.
\ea
For the theoretical predictions we shall take ${\cal{R}}(s)$ to be the NNLO CIPT/APT
CORGI result of Eq.(74). Starting with
${\tilde{\Lambda}}^{(5)}_{\overline{MS}}= 216 \rm{MeV}$ for ${N}_{f}=5$
, corresponding to the world average value ${\alpha}_{s}({M}_{Z})=0.1172$ \cite{r40a}, we
demand that ${\cal{R}}(s)$ remains continuous as we cross quark mass thresholds. This then
determines ${\tilde{\Lambda}}^{({N}_{f})}_{\overline{MS}}$ for ${N}_{f}=4,3,2$. We take standard
values for current quark masses for the light quarks \cite{r40a} : ${m}_{u}=3.0 \rm{MeV}$, ${m}_{d}=6.75 \rm{MeV}$,
${m}_{s}=117.5 \rm{MeV}$, and also from \cite{r40a}  we take the values for
pole masses of the heavy quarks ${m}_{c}=1.65 \rm{GeV}$, and ${m}_{b}=4.85
\rm{GeV}$. The approximate result \cite{r3} uses pole masses in Eq.(81), so
we use pole masses where we can. Using these
values for the quark masses and ${\alpha}_{s}({M}_{Z})$, we plot the resulting
${R}_{{e}^{+}{e}^{-}}(s)$ in Fig.6. The solid line corresponds to the CORGI APT
result for ${\cal{R}}(s)$ in Eq.(74). The dashed curve corresponds to the standard
NLO fixed-order CORGI result,
\be
{\cal{R}}_{CORGI}(s)={a}_{0}(s)+\left({X}_{2}-\frac{{\pi}^{2}{b}^{2}}{12}\right){a}_{0}^{3}(s)\;.
\ee
The standard fixed-order result breaks down at ${s}={\Lambda}_{D}^{2}=0.4114 \rm{GeV}^2$, where
there is a Landau pole. The APT result smoothly freezes in the infra-red. The dashed-dot
curve shows the parton model result (i.e. assuming ${\cal{R}}(s)=0$).\\

For a recent comprehensive review of the experimental data for ${R}_{{e}^{+}{e}^{-}}(s)$ at low energies
see Ref.\cite{whalley}.
The experimental data we have used comes from a variety of sources.
From the two pion threshold up to $\sqrt{s} = 1.43 \rm{GeV}$ we use references
\cite{r41h}, the data from these references is given as individual exclusive
channels which must be combined to obtain the full hadronic cross
section. In the region between
 $1.43 \rm{GeV}$ and $2.0 \rm{GeV}$ we use data from
\cite{r41a}, \cite{r41b}, references \cite{r41c}, \cite{r41d} are used in the
region between $2.0 \rm{GeV}$ and $ 5.0 \rm{GeV}$. From
 $5.0 \rm{GeV}$ to $7.25 \rm{GeV}$
we use \cite{r41e}, and from
 $7.25 \rm{GeV}$ to $10.52 \rm{GeV}$ we use
\cite{r41f}, \cite{r41g}. These sets of data all give the inclusive total
hadronic cross section. Above $10.52 \rm{GeV}$ we insert the NNLO CORGI APT
prediction for $R_{e^+e^-}$, this is represented by the continuous line in Fig.7.

\begin{figure}
\begin{center}
 \psfrag{a}[br][br]{\scriptsize$J/\Psi$\normalsize}
 \psfrag{b}[bl][bl]{\scriptsize$\Psi(2S)$\normalsize}
 \psfrag{c}[br][b]{\scriptsize$\Upsilon(1S)$\normalsize}
 \psfrag{d}[b][b]{\scriptsize$\Upsilon(2S)$\normalsize}
 \psfrag{e}[bl][br]{\scriptsize$\Upsilon(3S)$\normalsize}
 \psfrag{x}[t][bl]{$\sqrt{s}/\rm{GeV}$}%
 \psfrag{y}[tr][br]{$R_{e^+e^-}(s)$}

\epsfig{file=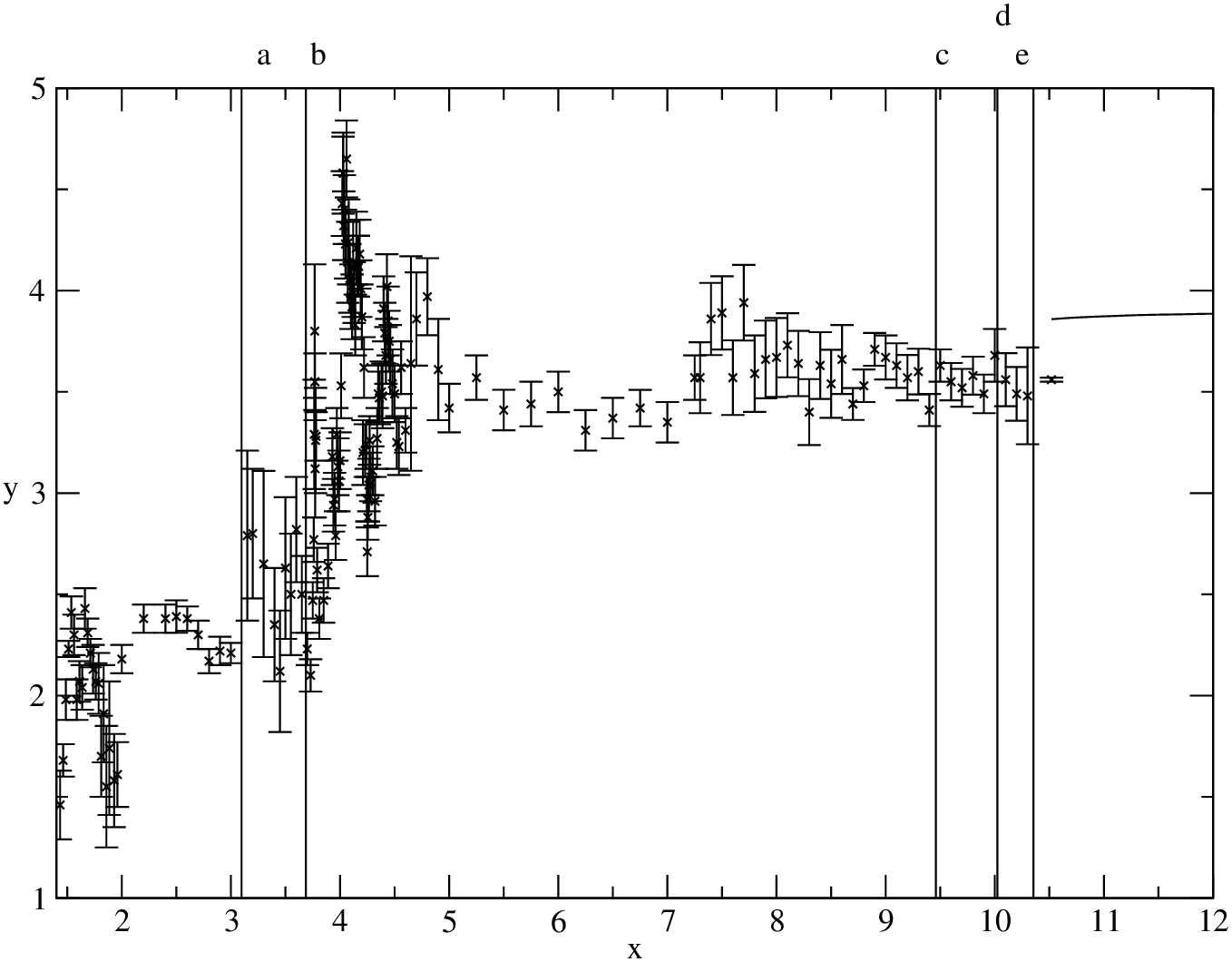,height=7.0cm,width=10.5cm}
\caption{Data used to compare with model, statistical errors shown only.}
\end{center}
\end{figure}

In order to simplify the analysis of the data we did not use overlapping
datasets, instead where one dataset overlapped another we simply took the
better, smaller error, dataset in the region of the overlap in
$\sqrt{s}$. Errors were dealt with by taking each data point and calculating
the effect of its statistical and its systematic error. The effect of its
statistical error was added in quadrature with the other statistical errors.
The contribution from the systematic error was added to the other
systematic errors from the same dataset, then the contribution from the
systematic errors of each dataset were added in quadrature with each other
and the contribution from the statistical errors.\\

We also need to consider the effect of narrow resonances not included in the
data, we employ the same approach as used in \cite{r12}. We assume that the
narrow resonances have a relativistic Breit-Wigner form
\be
R_{res}(s) = \frac{9}{\alpha^2}B_{ll}B_{h}\frac{M^2\Gamma^2}{(s-M^2)^2 +
  M^2\Gamma^2},
\ee
where $\alpha$ is the QED coupling, and $M,\Gamma,B_{ll},B_{h}$ are the mass,
width, lepton branching ratio, and hadron branching ratio respectively. We
are assuming a narrow resonance i.e. $\Gamma$ is small, so we approximate the
resonance with a delta function
\be
R_{res}(s) =
\frac{9}{\alpha^2}B_{ll}B_{h}{M\Gamma\pi}\frac{M\Gamma/\pi}{(s-M^2)^2 +
  M^2\Gamma^2} \approx \frac{9}{\alpha^2}B_{ll}B_{h}M\Gamma\pi\delta(s-M^2).
\ee

The compilation
of data for ${R}_{{e}^{+}{e}^{-}}$ is shown in Fig.7. Narrow resonances are indicated
by the vertical lines. Unfortunately it is not possible to directly compare the experimental
data with the theoretical predictions. This is because there is not a direct correspondence
between the quark mass thresholds in perturbation theory and the hadronic resonances. This
difficulty can be overcome if one employs a ``smearing procedure''. We shall employ the
method proposed by Poggio, Quinn and Weinberg \cite{r3}, defining the smeared quantity
\be
{\bar{R}}_{{e}^{+}{e}^{-}}(s;{\Delta})=\frac{\Delta}{\pi}\int_{0}^{\infty}{dt}\frac{{R}_{{e}^{+}{e}^{-}}(t)}
{{(t-s)}^{2}+{\Delta}^{2}}\;.
\ee
 ${R}_{{e}^{+}{e}^{-}}(s)$ itself is related to the vacuum-polarization function ${\Pi}(s)$ of Eq.(4) by,
 \be
 2i{R}_{{e}^{+}{e}^{-}}(s)=\Pi(s+i\epsilon)-\Pi(s-i\epsilon)\;,
 \ee
 that is it is the discontinuity across the cut. The smeared ${\bar{R}}_{{e}^{+}{e}^{-}}(s;\Delta)$
 can be written as
 \be
 2i{\bar{R}}_{{e}^{+}{e}^{-}}(s;\Delta)=\Pi(s+i\Delta)-\Pi(s-i\Delta)\;.
\ee
\begin{figure}
\begin{center}
  \psfrag{x}[t][bl]{$\sqrt{s}/\rm{GeV}$}
  \psfrag{y}[tr][br]{$\overline{R}(s;\Delta)$}

\epsfig{file=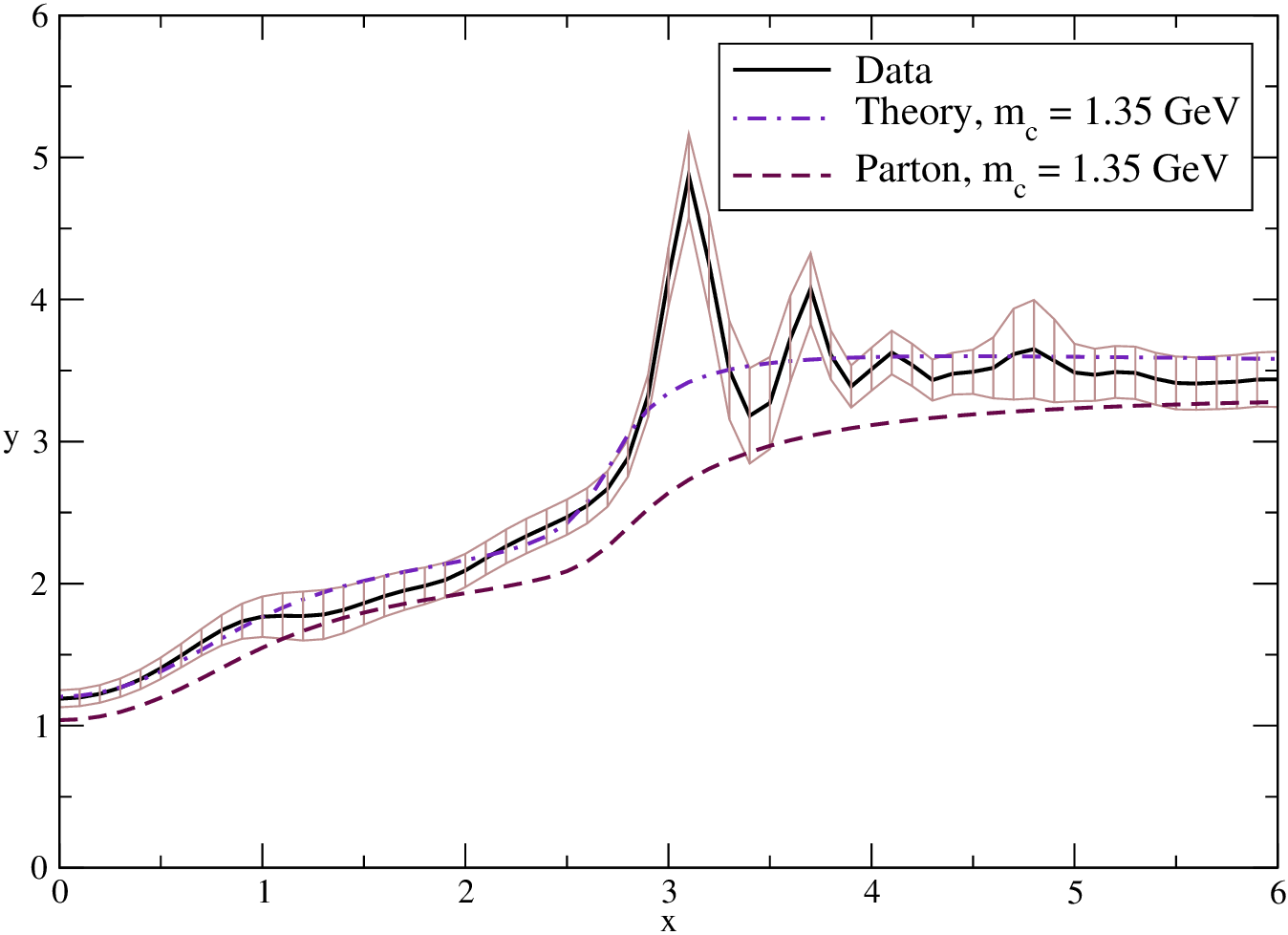,height=7.0cm,width=10.5cm}
\caption{(a): $\overline{R}(s;\Delta)$ in the charm region, with $\Delta = 1 \rm{GeV}^2$.}
\end{center}
\end{figure}
\setcounter{figure}{7}
If $\Delta$ is sufficiently large one is kept away from the cut, and is
insensitive to the infrared singularities which occur there. If both data
and theory are smeared they can then be compared. In this way one hopes to
minimise the contribution of the ${\cal{R}}_{NP}$ component.
One needs to choose
$\Delta$ sufficiently large that resonances are averaged out. For the
charm region it turns out that $\Delta=3{\rm{GeV}}^{2}$ is a good choice,
whilst for lower energies $\Delta=1{\rm{GeV}}^{2}$ is adequate. In Fig.8(a)
we choose $\Delta=1{\rm{GeV}}^{2}$. ${\bar{R}}_{{e}^{+}{e}^{-}}(s;\Delta)$
obtained from the data is represented by the solid line. The dashed-dot line is
the smeared NNLO CORGI APT prediction, assuming the quark mass thresholds as
above with the exception of the charm quark whose mass is taken to be ${m}_{c}=1.35 \rm{GeV}$
for reasons which we shall shortly discuss.
The dashed line is the parton model prediction. The shaded region denotes the
error in the data. It is clear that in the charm region the averaging is insufficient,
although for lower energies the agreement is extremely good. In Fig.8(b) we show
the corresponding plot with ${\Delta}=3 {\rm{GeV}}^{2}$. There is now good agreement
between smeared theory and experiment over the whole $s$ range, for $m_c =
1.35 \rm{GeV}$.
\begin{figure}
\begin{center}
  \psfrag{x}[t][bl]{$\sqrt{s}/\rm{GeV}$}
  \psfrag{y}[tr][br]{$\overline{R}(s;\Delta)$}

\epsfig{file=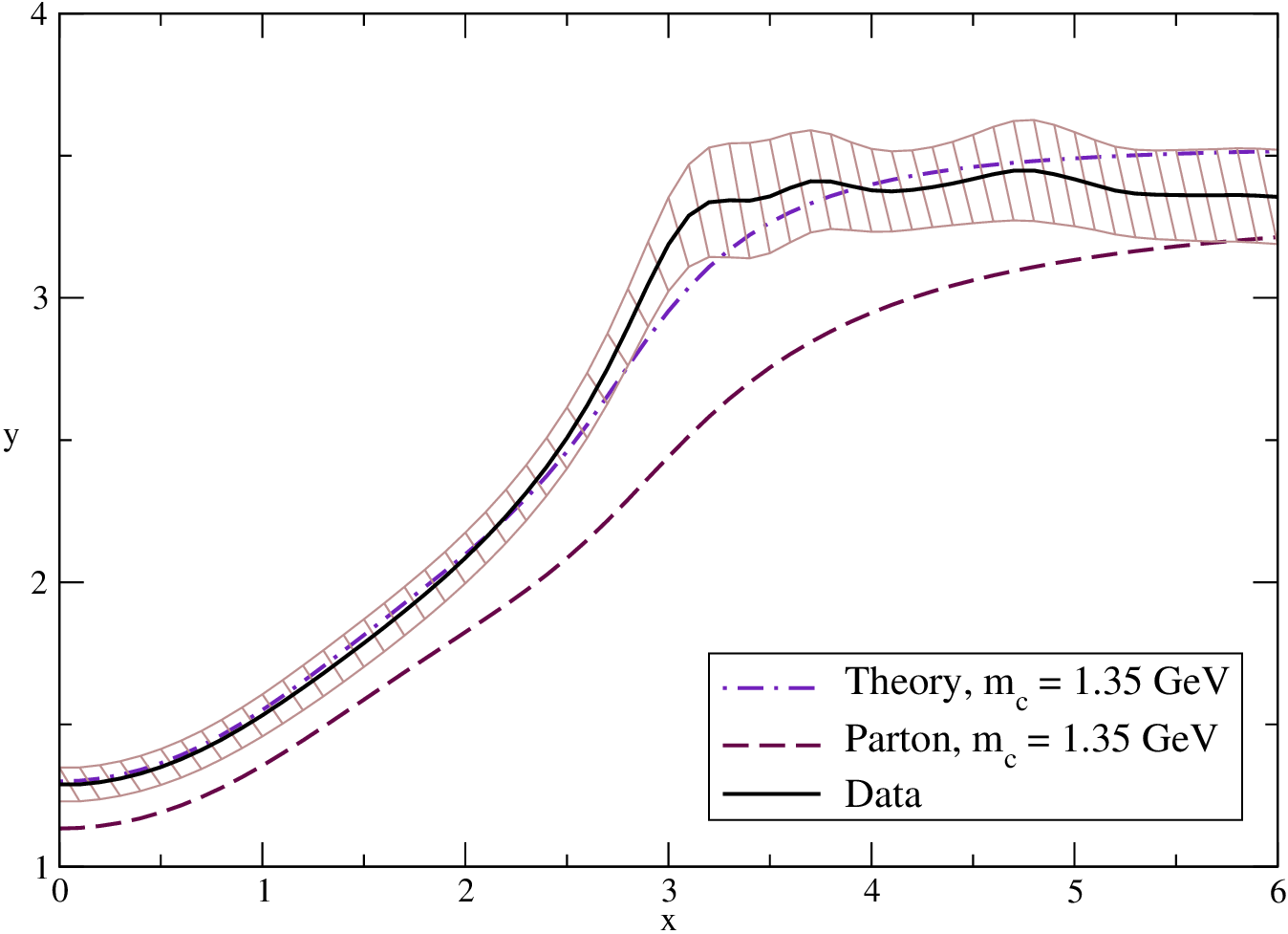,height=7.0cm,width=10.5cm}
\caption{(b): $\overline{R}(s;\Delta)$ in the charm region, with $\Delta = 3 \rm{GeV}^2$.}
\end{center}
\end{figure}
\setcounter{figure}{7}
Whilst we have indicated an error band associated with the data, we have not indicated an error band for
the theory prediction. There are several potential sources of error to consider.
The first is the choice of renormalisation scale. Our viewpoint would be that the
use of the CORGI scale corresponds to a complete resummation of ultraviolet
logarithms, which in the process results in a cancellation of $\mu$-dependent
logarithms contained in the coupling $a({\mu}^{2})$ and in the perturbative coefficients.
As we have argued elsewhere \cite{r39} attempts to estimate a theoretical
error on the perturbative predictions by making {\it ad hoc} changes in the
renormalization scale are simply misleading, and give no information on the importance
of uncalculated higher-order corrections.
A common approach, for instance, is to use scales ${\mu}^{2}=xs$ where $x$
is varied between $x=\frac{1}{2}$ and $x=2$, with $x=1$ providing a central value.
We should note, however, that were we to have used such a procedure it would
not have led to a noticeable difference in the theory curves, since the APT has
greatly reduced scale-dependence, as has been noted elsewhere \cite{apt4}. A more
important uncertainty is the precise value of the quark masses assumed, and in
particular the choice of the charm quark mass $m_c$. To illustrate how this effects
the results we show in Fig.8(c) the curves obtained if we assume ${m}_{c}=1.65\rm{GeV}$.
As can be seen the theory curve is now inconsistent with the data in the charm
region, although for lower energies where the charm quark has decoupled, the
agreement is again good.\\
\begin{figure}
\begin{center}
  \psfrag{x}[t][bl]{$\sqrt{s}/\rm{GeV}$}
  \psfrag{y}[tr][br]{$\overline{R}(s;\Delta)$}

\epsfig{file=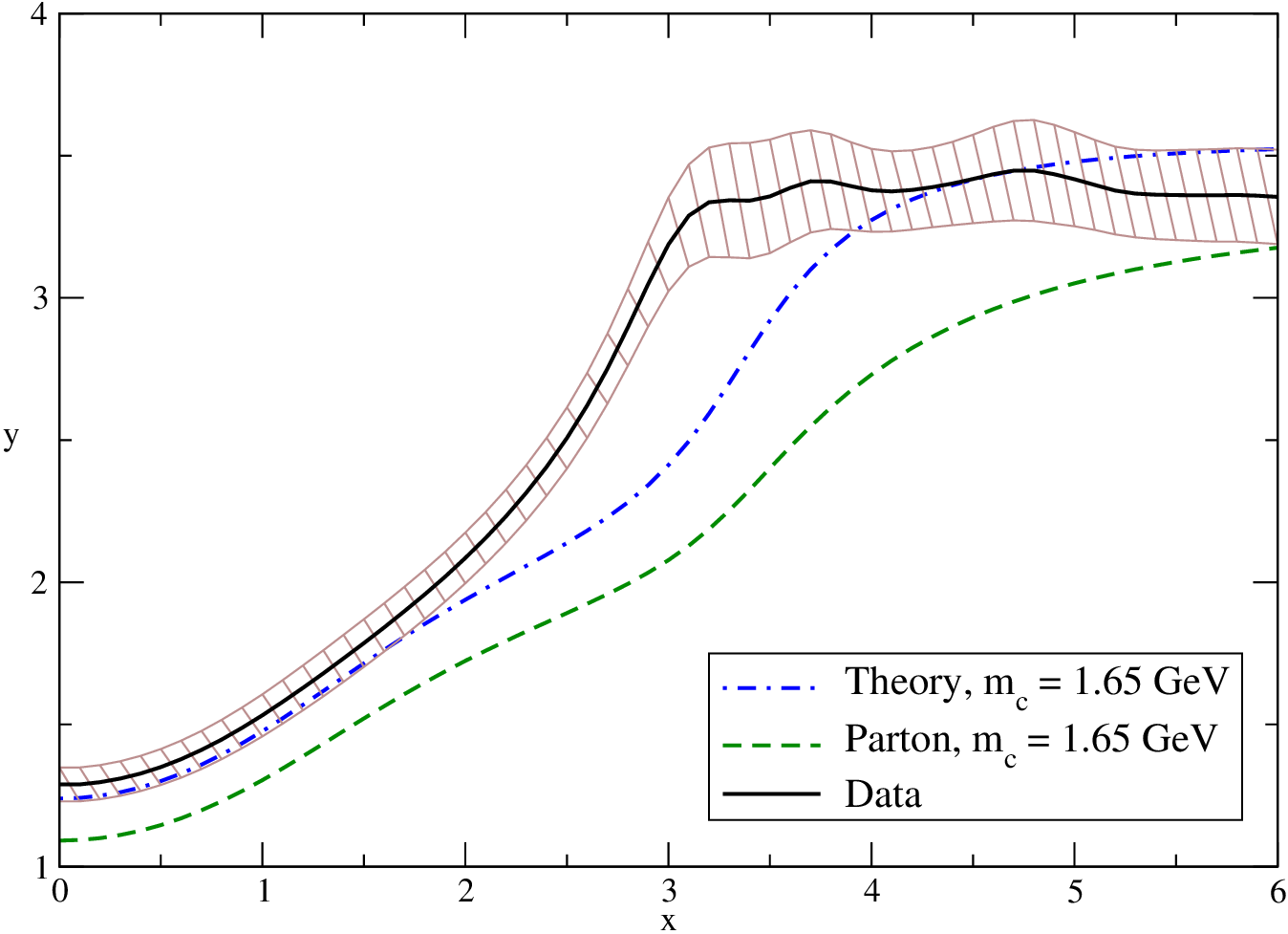,height=7.0cm,width=10.5cm}
\caption{(c): $\overline{R}(s;\Delta)$ in the charm region, with $\Delta = 3
  \rm{GeV}^2$ here $m_c = 1.65 \rm{GeV}.$}
\end{center}
\end{figure}
The uncertainty in the mass of the charm quark is
exceptionally large. Looking at the different references used in \cite{r40a} a value $m_c = 1.35 \rm{GeV}$ for the pole mass is
reasonable,
and agrees well with \cite{r40b} which is referenced in \cite{r40a}. Part of the problem is the relationship between the
pole mass and the $\overline{MS}$ mass for the charm quark, where the
$\alpha_s^3$ contribution is larger than the $\alpha_s^2$
contribution. Obtaining the pole mass through $\overline{MS}$ mass
calculations, which is done in \cite{r40a}, is not very
satisfactory. Reference \cite{r40b}, which also fits low-energy ${R}_{{e}^{+}{e}^{-}}$ data,
 gives a pole mass of $m_c = 1.33-1.4
\rm{GeV}$, and so the choice of $1.35 \rm{GeV}$ is reasonable. \\

It is possible to extend the smearing to spacelike values
of $s$. We give the corresponding curves for ${\bar{R}}_{{e}^{+}{e}^{-}}(s;\Delta)$, with ${m}_{c}=1.35{\rm{GeV}}$,
over the range $-3<s<1$ ${\rm{GeV}}^{2}$ in Figs.9(a),9(b), for $\Delta=1{\rm{GeV}}^{2}$,
and $\Delta=3{\rm{GeV}}^{2}$, respectively. The agreement between theory and experiment
is extremely good in both cases.
\begin{figure}
\begin{center}
 \psfrag{x}[t][bl]{${s}/\rm{GeV}^{2}$}%
 \psfrag{y}[tr][br]{$\overline{R}(s;\Delta)$}

\epsfig{file=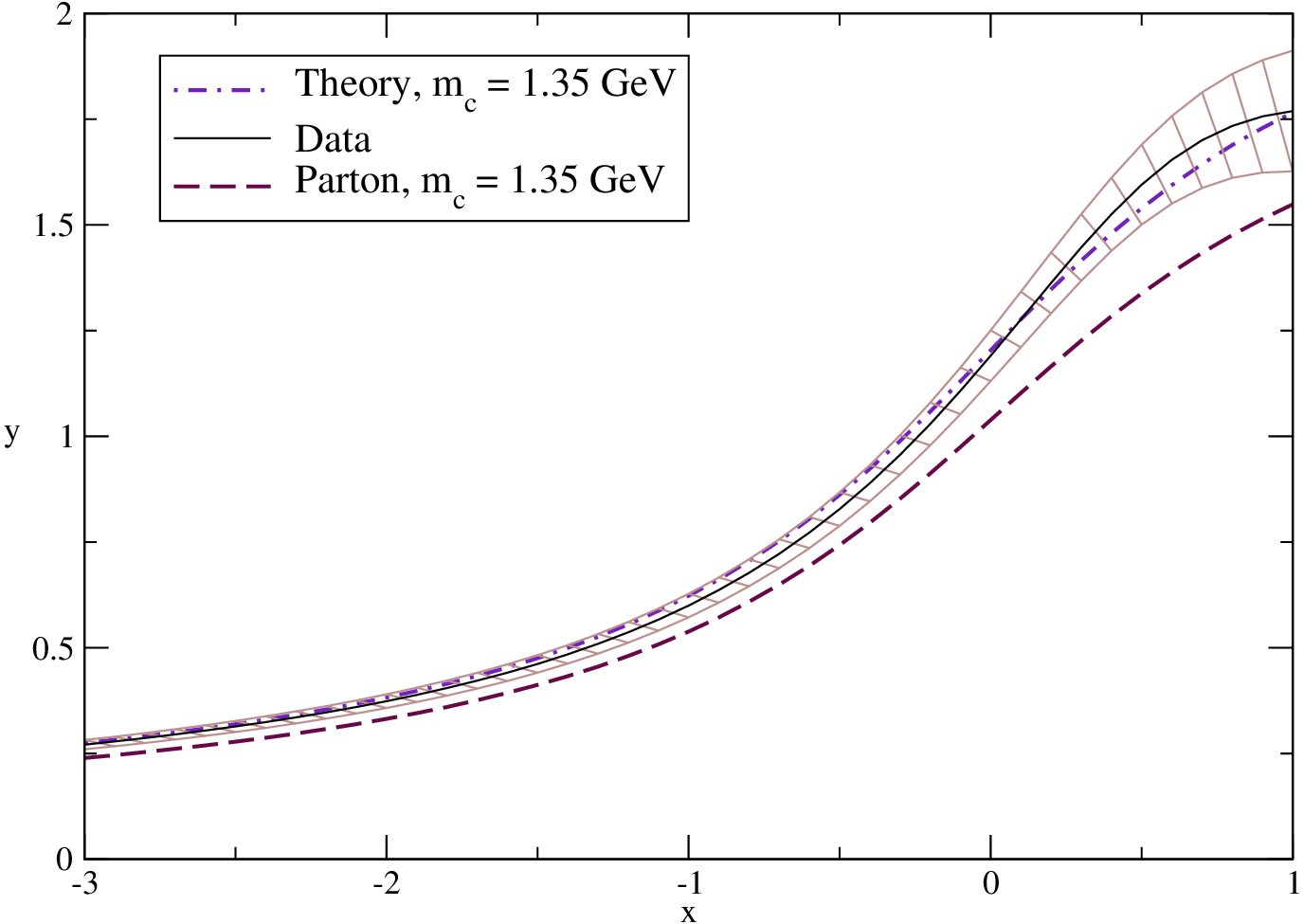,height=7.0cm,width=10.5cm}
\caption{(a): $\overline{R}(s;\Delta)$ in the spacelike region, with $\Delta = 1 \rm{GeV}^2$.}
\end{center}
\end{figure}
\setcounter{figure}{8}
\begin{figure}
\begin{center}
 \psfrag{x}[t][bl]{${s}/\rm{GeV}^{2}$}%
 \psfrag{y}[tr][br]{$\overline{R}(s;\Delta)$}

\epsfig{file=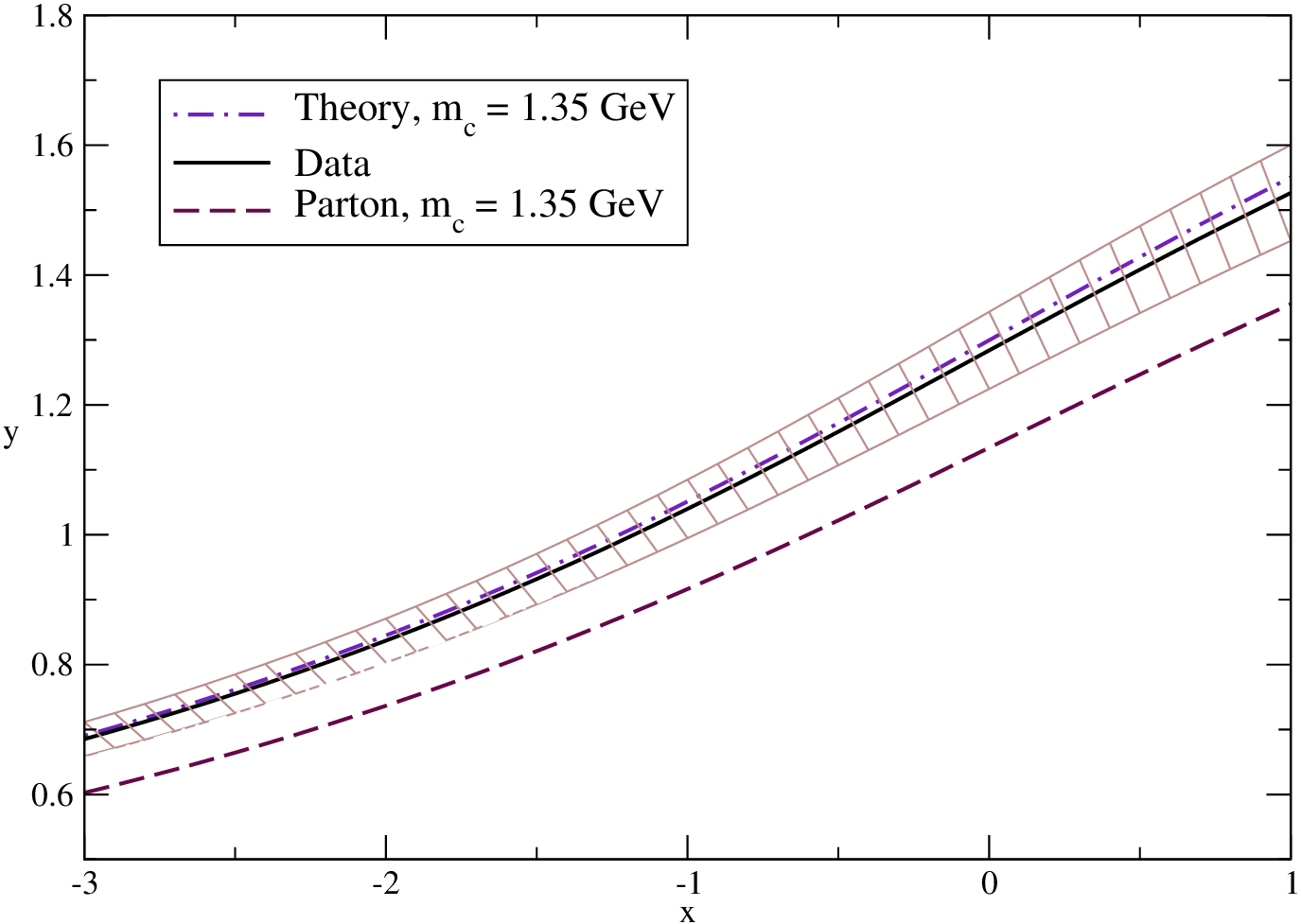,height=7.0cm,width=10.5cm}
\caption{(b):  $\overline{R}(s;\Delta)$ in the spacelike region, with $\Delta = 3 \rm{GeV}^2$.}
\end{center}
\end{figure}

\begin{figure}
\begin{center}
 \psfrag{x}[t][bl]{$\sqrt{s}/\rm{GeV}$}%
 \psfrag{y}[tr][br]{$\overline{R}(s;\Delta)$}

\epsfig{file=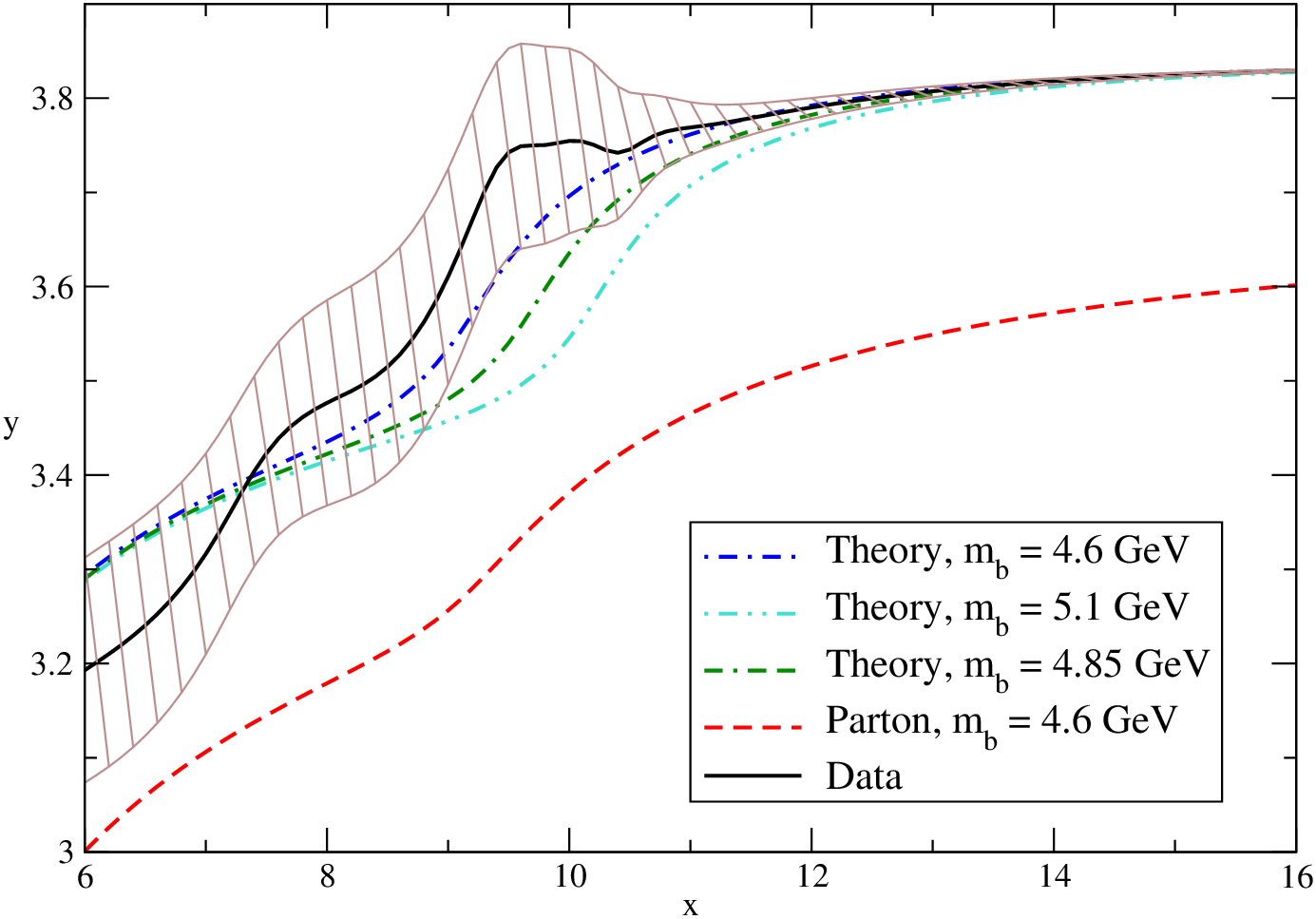,height=7.0cm,width=10.5cm}
\caption{$\overline{R}(s;\Delta)$ in the upsilon region, with $\Delta = 10 \rm{GeV}^2$.}
\end{center}
\end{figure}
In Fig.10 we show ${\bar{R}}_{{e}^{+}{e}^{-}}(s;\Delta)$ in
the upsilon region. The choice $\Delta=10{\rm{GeV}}^{2}$ works quite well, we show the
theory predictions for different $m_b$ values. A direct comparison between theory and
data which does not involve smearing is possible if one evaluates the area under the
${R}_{{e}^{+}{e}^{-}}(s)$ data, that is evaluates the integral,
\be
I(s)\equiv\int_{4{m}_{\pi}^{2}}^{s}{R}_{{e}^{+}{e}^{-}}(t){dt}\;,
\ee
where $s$ lies well above the low-energy resonances in the continuum. We show the theory
and experimental $I(s)$ over the range $5<\sqrt{s}<9 \rm{GeV}$ in Fig.11. There is extremely
good agreement. Finally we can avoid smearing by transforming ${R}_{{e}^{+}{e}^{-}}(s)$ to
obtain $D({Q}^{2})$ in the Euclidean region, using the dispersion relation of Eq.(75)

\begin{figure}
\begin{center}
 \psfrag{x}[t][bl]{$\sqrt{s}/\rm{GeV}$}
 \psfrag{y}[tr][br]{$I(s)$}

\epsfig{file=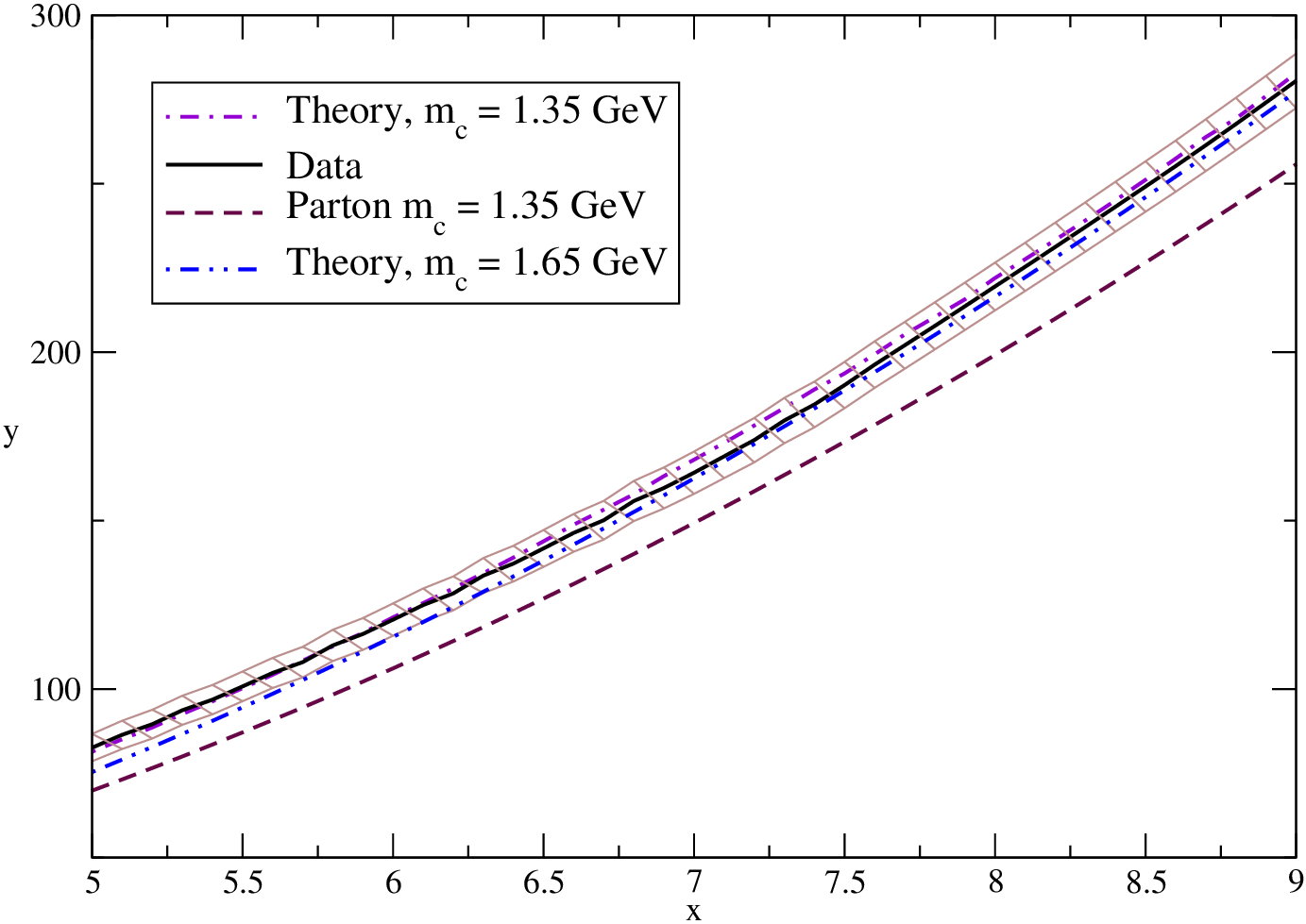,height=7.0cm,width=10.5cm}
\caption{Area under $R_{e^+e^-}(s)$}
\end{center}
\end{figure}

\be
D({Q}^{2})={Q}^{2}\int_{4{m}_{\pi}^{2}}^{\infty}\frac{ds}{{(s+{Q}^{2})}^{2}}{R}_{{e}^{+}{e}^{-}}(s)\;.
\ee
In practice we cannot integrate up to infinity so we just take the
sufficiently large upper limit of $10^6\rm{GeV}^2$. As noted earlier above $\sqrt{s}=10.52 \rm{GeV}$ the
NNLO CORGI APT prediction is used for the data.
 The theory and data results are shown in Figs.12,13. There
is good agreement. Our results are comparable to the fit obtained in
\cite{r42}, and to the results in \cite{r43}. We should also note that
very similar plots and fits to those we have presented in this Section
are included in Ref.\cite{apt3}, which uses instead the so-called Variational
Perturbation Theory (VPT) approach \cite{vpt1}.

\begin{figure}
\begin{center}
 \psfrag{x}[t][bl]{$Q/\rm{GeV}$}
 \psfrag{y}[tr][br]{$D(Q^2)$}

\epsfig{file=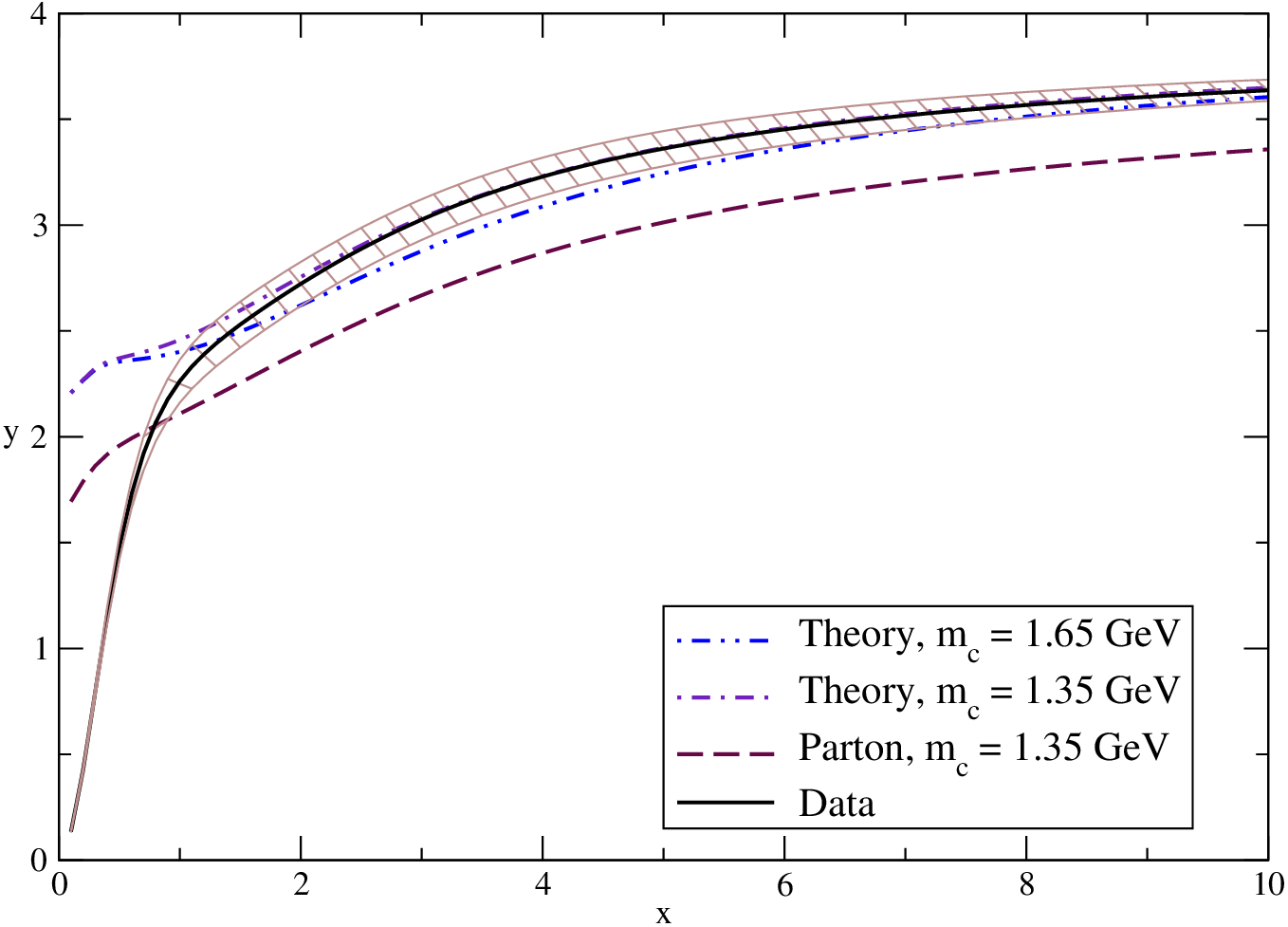,height=7.0cm,width=10.5cm}
\caption{$D(Q^2)$ calculated using APT.}
\end{center}
\end{figure}
\begin{figure}
\begin{center}
 \psfrag{x}[t][bl]{$Q/\rm{GeV}$}
 \psfrag{y}[tr][br]{$D(Q^2)$}

\epsfig{file=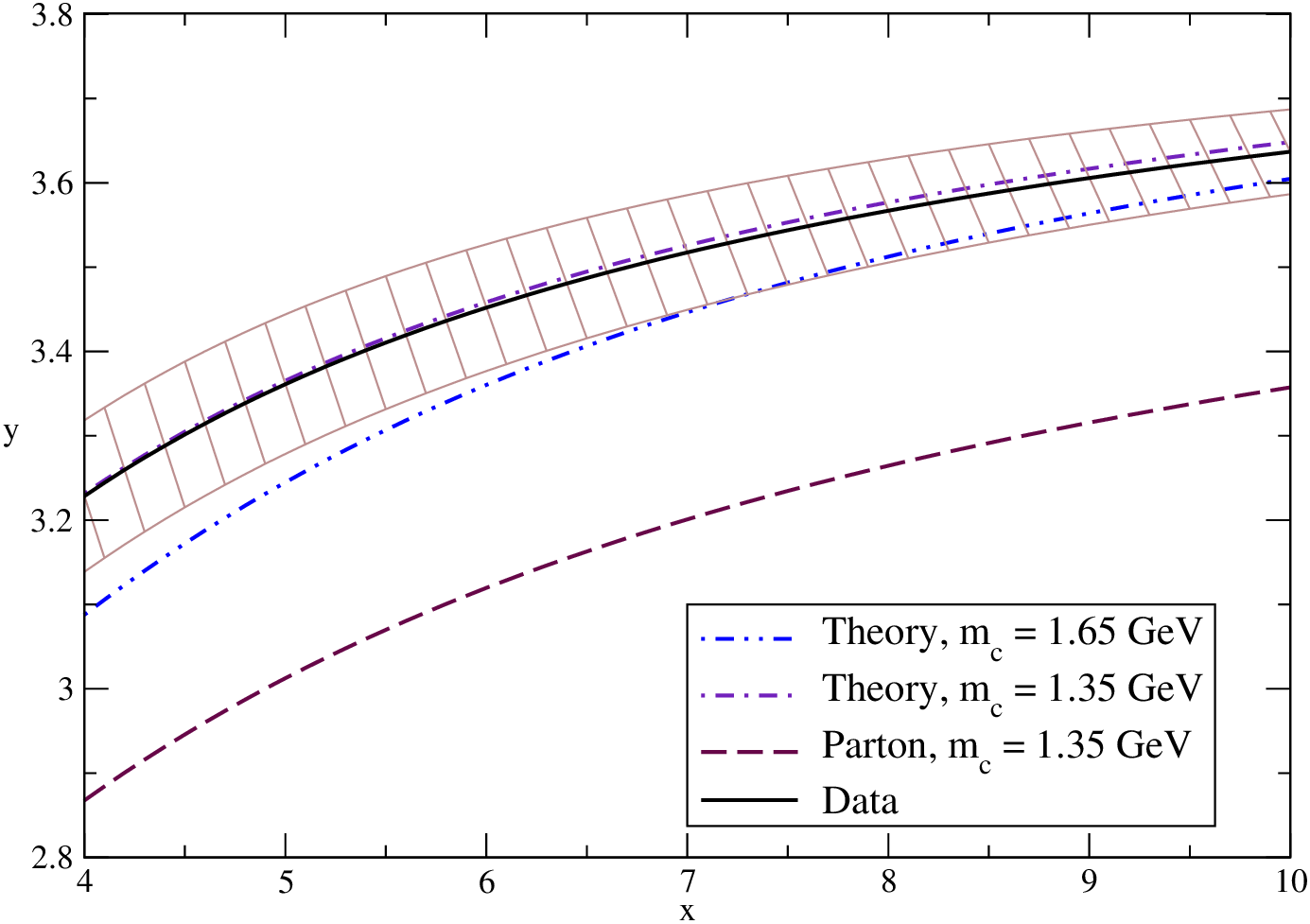,height=7.0cm,width=10.5cm}
\caption{Same as figure 12 but viewed over a smaller range.}
\end{center}
\end{figure}

\section*{5 Discussion and Conclusions}
The Analytic Perturbation Theory (APT) approach advocates the ``analytization''
of the terms in standard perturbation theory so that the perturbative expansion
is recast as an expansion in a basis of functions that have desirable analytic
properties, in particular the absence of unphysical ``Landau poles'' in $Q^2$ \cite{apt1}.
The functions in the Euclidean and Minkowski regions are interrelated
by the integral transforms of Eq.(7) (${\cal{D}}\rightarrow{\cal{R}}$) and Eq.(75)
(${\cal{R}}\rightarrow{\cal{D}}$). In a previous paper we pointed out the Minkowskian
formulation of APT for the quantity ${R}_{{e}^{+}{e}^{-}}$ was equivalent to the
all-orders resummation of a convergent subset of analytical continuation terms
\cite{r5}. This reorganisation of fixed-order perturbation theory gives apparent infrared
freezing to the limit $2/b$ to all-orders in perturbation theory, and the functions
${A}_{n}(s)$ at two-loop level could be written in closed form in terms of the
Lambert $W$ function. However, one might question whether this all-orders
perturbative freezing has any physical relevance. It is well-known that all-orders
perturbation theory by itself is insufficient, and that it must be complemented
by the non-perturbative Operator Product Expansion (OPE) \cite{r1,r2}. It is clear
that the OPE breaks down as $s\rightarrow{0}$, since it is an expansion in powers of
${\tilde{\Lambda}}^{2}/s$. In this paper we have shown how both the PT and the
OPE components can remain defined in the infrared limit. Writing a Borel representation
for the PT component one finds that it is ambiguous because of the presence of singularities
on the integration contour, termed infrared renormalons {\cite{r1}}. These ambiguities,
however, are of precisely the same form as OPE terms, and a regulation of the
singularities in the Borel integrand induces a definition of the OPE coefficients,
allowing the two components to be defined. We showed that the Borel integral
representation inevitably breaks down at a critical energy ${s}_{L}$ which we
referred to as the ``Landau
 divergence''. For Minkowskian quantities the Borel
Transform contains an oscillatory factor which means that the Borel integral
remains defined at $s={s}_{L}$. For $s<{s}_{L}$ one needs to switch to an
alternative Borel representation, which has ambiguities due to ultraviolet
renormalon singularities on the integration contour. Correspondingly the OPE
should be resummed and recast in the form of an expansion in powers of
 ${s}/{\tilde{\Lambda}}^{2}$. The UV renormalon ambiguities in the Borel
integral are then of the same form as the terms in the modified OPE, and regulating
the modified Borel integral induces a definition of the coefficients in the
modified OPE, allowing both components to be defined. The modified Borel integral
freezes to $2/b$ in the infrared thanks to the presence of the oscillatory factor, whilst
the modified OPE component will also contribute to the infrared freezing behaviour
since resummation of the standard OPE can result in $s$-independent terms which
can give a nonzero freezing limit, as in the toy example of Eq.(51). As we noted
we did not expect to be able to determine the infrared behaviour from perturbation
theory alone, but the existence of a perturbative component which can be defined
using a reorganised version of fixed-order perturbation theory at all energies is
important. In particular the perturbative component dominates in the ultraviolet and
may possibly provide a good approximation into the low-energy region.
We explicitly constructed the all-orders Borel representations using the all-orders
leading-$b$ approximation for ${\cal{R}}(s)$ \cite{r29a}, and a one-loop coupling.
We could express the Borel integral in closed form in terms of exponential integral
functions (Eq.(47)). With the standard continuation of the $Ei(n,w)$ functions defined
by Eq.(44) the result for ${\cal{R}}^{(L)}_{PT}(s)$ of Eq.(47) is a function of $s$
which is well-defined at all energies, freezing to $2/b$ in the infrared, and continuous
at $s={s}_{L}$. The two-loop Borel representation was also discussed. The details
are similar to the one-loop case, with a modified oscillatory factor and a shifted
value of $s_L$, the modified Borel representation again freezes to $2/b$ in the
infrared. At both one-loop and two-loops the APT modification of fixed-order
perturbation theory corresponds to keeping the oscillatory factor in the Borel
integrand intact, and expanding the remainder. As a result the APT results
should be asymptotic to the Borel representations at all energies, underwriting
the validity of the all-orders perturbative freezing behaviour. It should be
noted that we have somewhat oversimplified our discussion of the OPE contribution.
The OPE coefficients are not constant, as in the toy example of Eq.(51), but
are functions of $a$, ${\cal{C}}_{n}(a)$. Each coefficient will
 involve a perturbation series in $a$ which is divergent with $n!$ growth of coefficients,
and can be defined using a Borel representation. As defined by analytic continuation from
the OPE for ${\cal{D}}_{NP}$ to that for ${\cal{R}}_{NP}$, the corresponding Borel
integrands will contain the same oscillatory factors, enabling ${\cal{C}}_{n}(a)$ to
remain defined at $s={s}_{L}$, and for $s<{s}_{L}$ one switches to the modified
Borel representation. We should note that the difficulty of uniquely extending
the Borel representation for Minkowskian quantities into the infrared has also
been discussed in Ref.\cite{r44}, but with differing conclusions to us. A
more closely related discussion concerning the significance and interpretation
of the Landau Pole is given in Ref.\cite{r31}. The modified Borel representation
of Eq.(50) and the promotion of UV renormalon singularities to the positive axis
in the Borel $z$-plane has also been discussed in Ref.\cite{r33}.

 Whilst the Minkowskian version of APT is underwritten by
a Borel representation valid at all energies, this is not the case for the Euclidean
version. There is no oscillatory factor in the integrand in the Euclidean case,
and the Borel integral will potentially diverge as one approaches ${s}_{L}$. However,
we showed that working in leading-$b$ approximation ${\cal{D}}_{PT}$ was finite at
${s}_{L}$ thanks to a cancellation between the infinite set of IR and UV renormalon
residues. For individual renormalon singularities the Borel integral is divergent.
By switching to the modified Borel representation one can then define a ${\cal{D}}_{PT}$
component which in fact freezes to zero in the infrared. This is interesting and similar
perturbative freezing is also found for structure function sum rules \cite{pb}. The key
point, however, is that no analogue of the Minkowskian APT reorganisation of fixed-order
perturbation theory is possible in the Euclidean case, and one is restricted to the
leading-$b$ approximation in exhibiting the perturbative freezing.

In the final Section we performed fits of NNLO APT results to low energy ${R}_{{e}^{+}{e}^{-}}$
data. We needed to introduce quark masses approximately, and in order to avoid
ambiguities due to the precise location of quark mass thresholds, and to minimise the contribution
of the ${\cal{R}}_{NP}$ component,
 we used a smearing
procedure. Extremely good agreement between theory and data was found. \\

An obvious further application would be to use the APT approach in the analysis of
the tau decay ratio
 and in particular the estimation of the
uncertainty in ${\alpha}_{s}(M_Z)$ which such measurements imply \cite{r7,apt4}. In Ref.\cite{r7}
this was estimated by comparing NNLO CIPT in the CORGI approach, with an
all-orders resummation based on the leading-$b$ result. However, in fact CIPT
for the tau decay ratio is {\it not} equivalent to the APT approach and corresponds
to an expansion in a different basis of functions. In particular the resulting functions
are  {\it discontinuous} at $s={s}_{L}$. We hope to study this further in a
future publication.

\section*{Acknowledgements}
We thank Andrei Kataev and Paul Stevenson for entertaining discussions on
infrared freezing in perturbative QCD.
D.M.H. gratefully acknowledges receipt of a PPARC UK Studentship.

\end{document}